\documentclass[structabstract]{aa}  
\usepackage{graphicx}
\usepackage{txfonts}

\usepackage{natbib}
\bibpunct{(}{)}{;}{a}{}{,} 

\newcommand{\ind}[1]{_{\mathrm{#1}}}

\begin{document}
   \title{Mode width fitting with a simple bayesian approach.\\Application to CoRoT targets HD~181420 and HD~49933\thanks{The CoRoT space mission, launched on
   2006 December 27, was developed and is operated by the CNES, with
   participation of the Science Programs of ESA, ESA's RSSD, Austria,
   Belgium, Brazil, Germany and Spain.}}
\author{Patrick Gaulme
          \and
          Thierry Appourchaux
          \and Patrick Boumier}
         
\offprints{P. Gaulme}

   \institute{Institut d'Astrophysique Spatiale (UMR 8617), Facult\'e des Sciences d'Orsay, Universit\'e Paris Sud, B\^atiment 121, F-91405 ORSAY Cedex  \\
              \email{Patrick.Gaulme@ias.u-psud.fr}
        }

\titlerunning{Mode width fitting with a simple Bayesian approach}
\authorrunning {Gaulme et al.}
\abstract
 {}
{We investigate the asteroseismology of two solar-like targets as observed with the CoRoT satellite, with particular  attention paid to the mode fitting. HD~181420 and HD~49933 are typical CoRoT solar-like targets (156 and 60-day runs). The low signal-to-noise ratio (SNR) of about $3-10$ prevents us from unambiguously identifying the individual oscillation modes. In particular, convergence problems appear at the edges of the oscillation spectrum.}
{We apply a Bayesian approach to the analysis of these data. We compare the global fitting of the power spectra of this time series, obtained by the classical maximum likelihood (MLE) and the maximum a posteriori (MAP) estimators.}
{We examine the impact of the choice of the priors upon the fitted parameters. We also propose to reduce the number of free parameters in the fitting, by replacing the individual estimate of mode height associated with each overtone by a continuous function of frequency (Gaussian profile). }
{The MAP appears as a powerful tool to constrain the global fits, but it must be used carefully and only with reliable priors. The mode width of the stars increases with the frequency over all the oscillation spectrum. }
\keywords{statistics --
                detection --
                data analysis -- asteroseismology
               }

\maketitle

\section{Introduction}

The CNES CoRoT satellite has been orbiting the Earth at an altitude of 800 km, in a polar orbit, since late December 2006. Its scientific goals are divided into two themes: asteroseismology and the detection of transiting exo-planets. In this paper, we focus on the analysis of seismological data, in particular on the analysis of the power spectra of the solar-like targets. The main parameters extracted from observations, in order to constrain stellar models, are the global mode frequencies, heights, widths, and parameters such as the rotational frequency splitting and the star inclination with respect to the line of sight. These parameters are extracted by fitting the power spectrum with a mathematical model, which consists of a sum of Lorentzian functions, justified by assuming that the oscillations are stochastically excited \citep{Duvall_harvey_86}. 

Such a technique has been commonly used with helioseismic measurements such as GONG, BiSON, GOLF, LOI, MDI (for a review of these instruments see \citealt{Howe_09}). However, thanks to the excellent signal to noise ratio (SNR) of solar data (40 with LOI, 300 with GOLF), the mode fitting could be done individually for each pair of modes (odd or even). Moreover, in the Sun's case, the inclination is well known as is the rotational splitting. The  difficulty of fitting a stellar power spectrum is firstly that the SNR is less than 10 (e.g. \citealt{Appourchaux_HD_08}) and secondly that both inclination and splitting are generally unknown, even if they have been roughly estimated with the help of complementary radial velocity measurements, for which the estimates depend strongly on the stellar model (size, mass).  

Therefore, in order to extract the maximum information from the solar-like stellar spectra obtained with CoRoT, it appeared unavoidable to globally fit the power spectra, otherwise, the global parameters such as the splitting, the inclination and the stellar background could not be properly fitted (Appourchaux et al., 2008). Consequently, the number of parameters to be simultaneously estimated increases substantially. As an example, in the case of the solar-like target HD 49933, where 14 overtones have been taken into account, a set of 75 free parameters was needed (see \citealt{Appourchaux_HD_08}). The fitted values are obtained with the {\it Maximum Likelihood Estimator} (MLE) (see Sect. 1). However, as already attempted with several CoRoT solar-like targets (HD 49933: \citealt{Appourchaux_HD_08}, HD 181420: \citealt{Barban_09}, HD 181906: \citealt{Garcia_09}), the MLE struggles to fit the power spectra when the SNR is lower than a given threshold (example in Sect. 1); the risk is to get a biased estimate of global parameters, because of the bad fitting of the ``external modes'' (at the edges of the power spectrum). 

In this context, \citet{Appourchaux_MAP_08} suggested adopting a Bayesian approach in order to make the most of all of the available {\it a priori} information we have on the star. Following those suggestions, \citet{Benomar_08} applied a Bayesian approach to fit the first run of HD 49933. The estimate of the parameters was determined by scanning the probability density function (PDF) associated with each of the 75 parameters with the numerical ``Monte-Carlo Markov Chain'' (MCMC) method. The result of this analysis gave a significantly more reliable result, particularly in preventing some demonstrably erroneous estimates (e.g. ``Dirac-like'' modes, see Sect. 1). The only real drawback of such a method is the long computational time (several days for a global fitting of HD 49933, Benomar, private communication). 

In this paper, we propose an {\it a minima} Bayesian approach, whose objective is to prevent erroneous solutions and to quickly extract the main trend of the stellar parameters. In that context, we use the {\it Maximum A Posteriori} (MAP) approach, which roughly consists of adding prior information to the MLE, then maximizing it in the same way. In other words, it is a regularized MLE. The main difference with respect to the MCMC method is that we do not explore the posterior PDF but simply determine the local maxima of the posterior PDF. The second purpose of the paper, which can be seen as a particular case of the MAP approach, is to reduce the total number of parameters to be fitted.
The work is organized as follows: Sect. 2 contains a summary of the MLE approach and gives an example of the typical problems that may appear; Sect 3 presents the Bayesian approach and discusses the risks we run by using it improperly; Sect. 4 is dedicated to the reduction of the number of free parameters and its application to the GOLF solar data; finally, Sect. 5 shows the application to 2 CoRoT solar-like targets: HD 181420 and HD 49933. We particularly pay attention to the resulting mode width dependence on the frequency. 

\begin{figure}
\includegraphics[width=8.4cm]{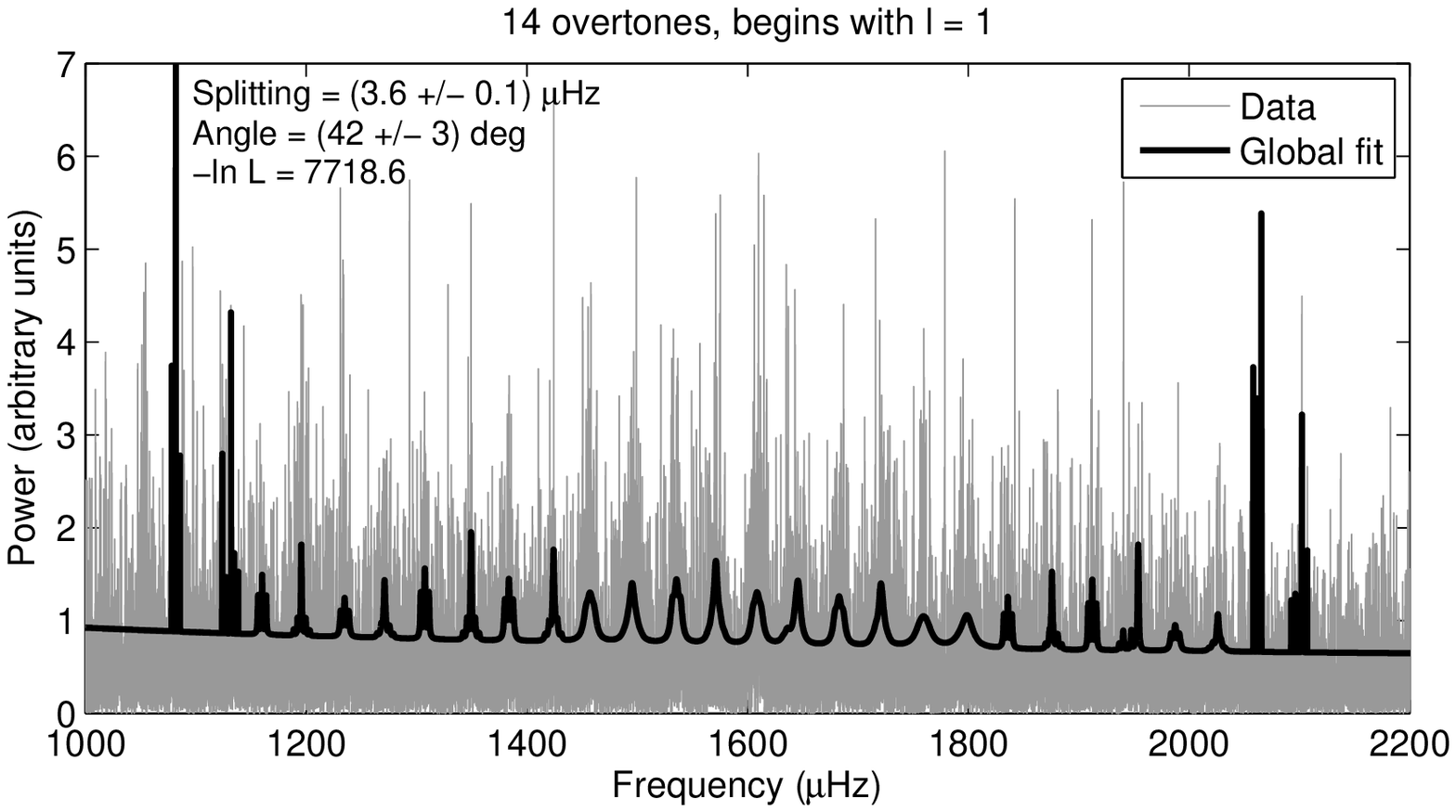}
\includegraphics[width=8.4cm]{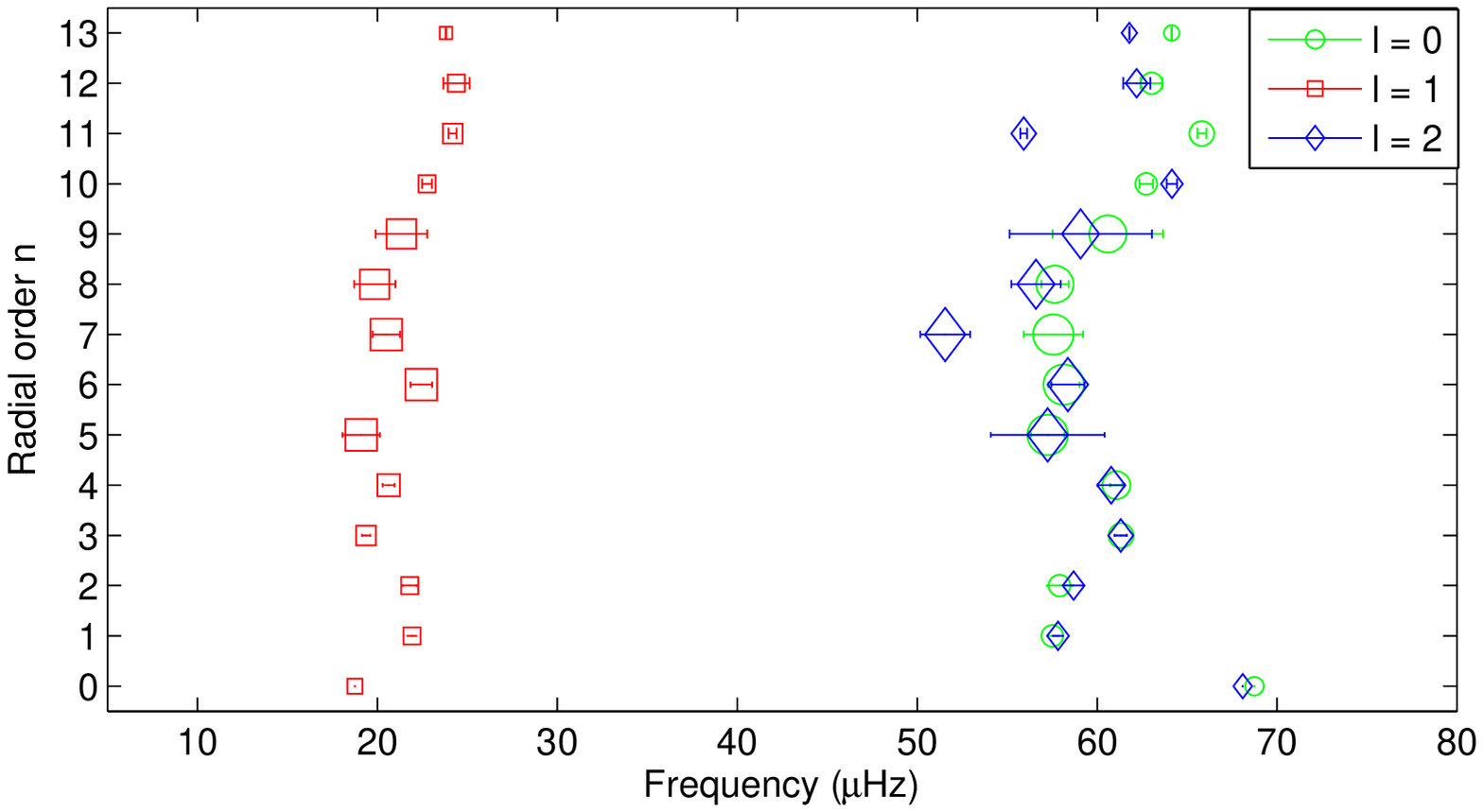}
\includegraphics[width=8.4cm]{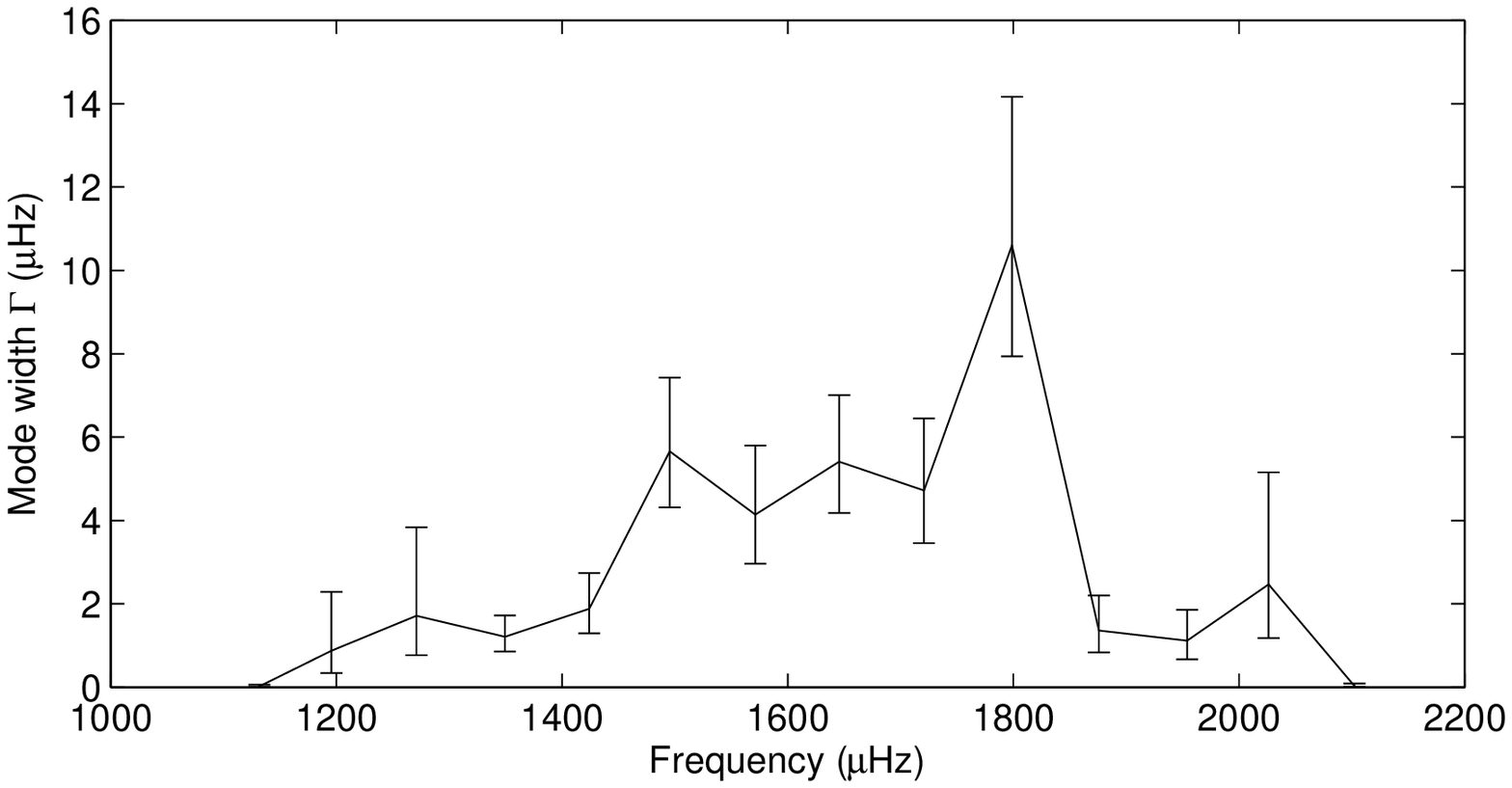}
\caption{Global fitting of the power spectrum of the solar-like target HD 181420 with the MLE. The fitting was realized over the frequency range $[600, 5000]\ \mu$Hz, by supposing 14 overtones and a single background profile. This fitting corresponds to scenario 1 in the Barban et al. (2009) paper, which is the most likely scenario according to the MLE estimator.  From top to bottom: 1) power spectrum and fitting, 2) an echelle diagram of the same, 3) estimate of the mode widths.}
\label{fig_MLE_181420}
\end{figure}

\section{Inability of MLE in the case of noisy spectra}
\subsection{Global power spectrum fitting}
With the assumption of stochastic excitation of the modes, the power spectrum of the eigenmodes is distributed around a mean profile following a $\chi^2$ with 2 degrees of freedom (d.o.f). It can be proved by considering the photometric time series as distributed  normally around its mean value. The power spectrum is the square of the modulus of the Fourier transform of the time series, that is the sum of the square of the real part $\int_0^\infty s(t)\cos(2\pi\nu t) \mathrm{d}t$ and the square of the imaginary part $\int_0^\infty s(t)\sin(2\pi\nu t) \mathrm{d}t$, which gives a  $\chi^2$ with 2 d.o.f. In other words, each point of the power spectrum is considered as a random variable which follows a $\chi^2$ distribution with 2 d.o.f.
Duvall \& Harvey (1986) showed that assuming that the excitation of the modes by turbulent convection is stochastic, the power spectrum must present a mean profile defined by a sum of Lorentzian functions as a function of the frequency: 
\begin{equation}
S_0\ =\ \sum_{n,\ell,m}\ \frac{H_{n,\ell}(i)}{\displaystyle1\ +\ 4\left(\frac{\nu - \nu_{n,\ell} - m \nu_S}{\Gamma^2_{n,\ell}}\right)^2} 
\label{S_0}
\end{equation}
where $\nu$ is the frequency, $H_{n,\ell}$, $\Gamma_{n,\ell}$ and $\nu_{n,\ell}$ the height, width and eigenfrequency (1 per pair $n,\ell$),  $\nu_S$ the splitting frequency and $i$ the inclination angle. The azimutal order $m$ is taken into account in the inclination angle dependence of the mode height \citep{Gizon_Solanki_03}.

The likelihood function is the probability of the data set D with respect to the model characterised by the parameter set $\lambda$. The best fit corresponds to the model for which the data set probability is maximum. By assuming each point of the power as independent, which is wrong if smoothed, $L = P(\rm{D}|\lambda)$ is:  
\begin{equation}
L\ =\ P(\rm{D}|\lambda,\rm{I})\ =\ \prod_i \frac{1}{S_0(\nu_i)}\ \exp\left(-{\displaystyle\frac{S_i}{S_0(\nu_i)}}\right)  
\end{equation}
where $S_i$ indicates the observed power spectrum value corresponding to the frequency bin $\nu_i$. In practice, it is equivalent and easier to minimize the quantity $l\ind{MLE}=-\log L$. We used the finite-differencing (or Quasi-Newton) minimization routine of MATLAB to perform our likelihood calculations; we also provided the analytic expression of the derivative to optimize the routine efficiency.  

Finally, the model function $S_0$ (Eq. \ref{S_0}) would be true if the power were only due to the stellar oscillations; actually, an additional term has to be added, in order to take into account the stellar noise associated with its surface granulation. This additional term may be written as a sum of a uniform noise level $B_0$ and a sum of background components as usually used in the helioseismology literature:
\begin{equation}
S_0'\ =\ S_0\ +\ \sum_{j=1}^3 \frac{K}{\displaystyle 1 + C \nu^p}\ +\ B_0.
\label{Eq_harvey}
\end{equation} 
The power term $p$, is generally fixed at 2 or 4. The SNR of a single mode is defined as the ratio of its height $H_{n,\ell}$ to the background $S_0' - S_0$. 

\subsection{When the MLE struggles to converge to a reasonable solution}
\label{sect_struggle}

Let us consider the fitting of the power spectrum of HD~181420, the first solar-like target to have been observed with CoRoT over a long run (156 days). The analysis of its power spectrum with the MLE has been carried out by Barban et al. (2009 hereafter B09), who included the results of several independent groups. In their work, they fitted the power spectrum by considering 14 overtones $n$ and degrees $\ell=[0,2]$.
Fitting 14 overtones until $\ellÊ=Ê2$ (Fig. \ref{fig_MLE_181420}) means fitting 75 parameters: 14 heights, 14 widths, 42 frequencies, a splitting, a stellar inclination, 2 stellar noise parameters (1 noise profile with a slope fixed at 2). 

In Fig. \ref{fig_MLE_181420} we present the global fitting of the power spectrum used by B09, obtained with the MLE for their scenario 1 (see also Sect. \ref{results}). The input parameters are close to the estimated values obtained in B09. It arises that when the signal-to-noise ratio of the power spectrum is too low, the MLE converges to a manifestly incorrect solution. In B09, the mode widths and heights of the external modes (lower and higher frequency) were rejected, in order to keep only the reliable fit values obtained with 10 overtones.

When the mode height becomes too low, the estimator tends to exactly fit some individual spikes of the power spectrum, that is, extremely high and narrow (very apparent in the frequency range [1100-1300] µHz). Hereafter we speak of ``Dirac-like'' convergence.
Such a behavior is not surprising since the MLE consists of maximizing the similarity between the data and the model. The Bayesian approach allows us to avoid such solutions.    

\section{The maximum a posteriori approach}
\subsection{Bayesian spirit in the sky}

Bayesian methods have been widely used in the deconvolution of astronomical images since the time of the restoration project for HST \citep{Bertero_95}, and also in cosmology (e.g. \citealt{Trotta_08}) in order to fully exploit available reliable prior knowledge. Bayesian reasoning often arises unconsciously: e.g. in the solar case, if the MLE estimator gives a value of the rotational splitting near 0.82 $\mu$Hz instead of $\simeq0.41\ \mu$Hz and an inclination of 0$^\circ$ instead of $\simeq90^\circ$, we reject that solution because we know it is wrong, not on the basis of a statistical criterion. The same occurred in B09 when eliminating the estimates of the external mode heights and widths. They seemed wrong although they were statistically correct, since they occurred at a local minimum of the parameter space.  

The Bayesian approach consists of using all of the prior information ``I'' that we have. Bayes theorem can be expressed as: 
\begin{equation}
P(\lambda | {\rm D,I})\ =\ \displaystyle{ {P(\lambda | {\rm I}) P({\rm D} | \lambda,{\rm I})} \over P({\rm D} | {\rm I})}.
\end{equation}
The Bayesian approach, compared with MLE, modifies the likelihood $P({\rm D} | \lambda,\rm{I})$ into a posterior distribution $P(\lambda | {\rm D,I})$ which takes into account the prior information $P(\lambda |{\rm I})$. The $P({\rm D | I})$ is a normalization term. In this paper we consider only the simplest way of applying the Bayesian approach, i.e. the maximum a posteriori estimator (MAP): 
\begin{equation}
\hat{\lambda}\ =\ \mbox{arg max}_\lambda\ P(\lambda | {\rm I})\ P({\rm D} | \lambda, {\rm I}). 
\label{Eq_MAP}
\end{equation}
Note the normalization term is not included and will not be taken into account in the following. Often, the prior is written as a Gaussian function centered around the expected value of the free parameter:  
\begin{equation}
P(\lambda | \rm{I})\  \propto \ \exp\left[-\frac{(\lambda - \lambda\ind{prior})^2}{ \sigma\ind{prior}^2}\right].
\end{equation}
Therefore, the function to be minimized can be expressed as:
\begin{equation}
l\ind{MAP}\ =\ l\ind{MLE}\ +\ \sum_\lambda \left(\frac{\lambda - \lambda\ind{prior}}{\sigma\ind{prior}}\right)^2 + \rm{C^{st}}.
\end{equation}
The MAP is the crudest Bayesian approach, but it allows us to improve the fits noticeably. 

\begin{figure}
\includegraphics[width=8.4cm]{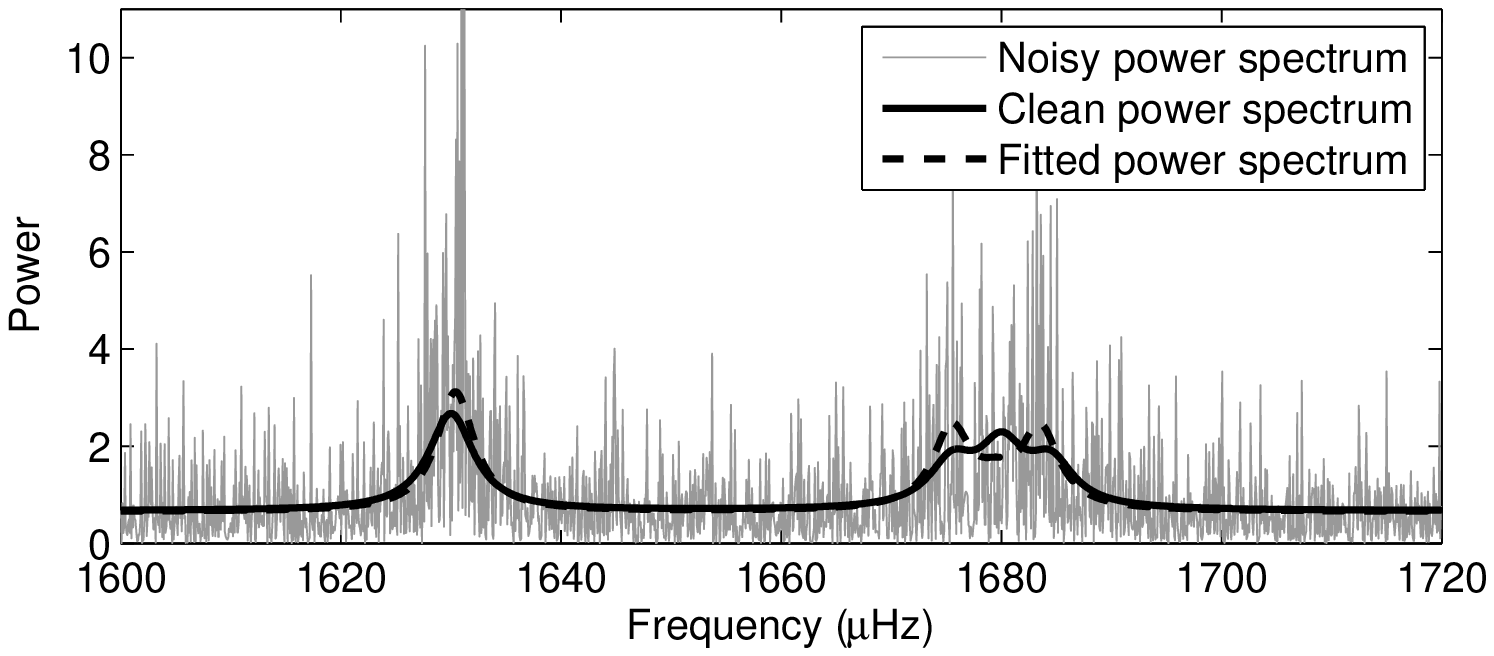}
\includegraphics[width=8.4cm]{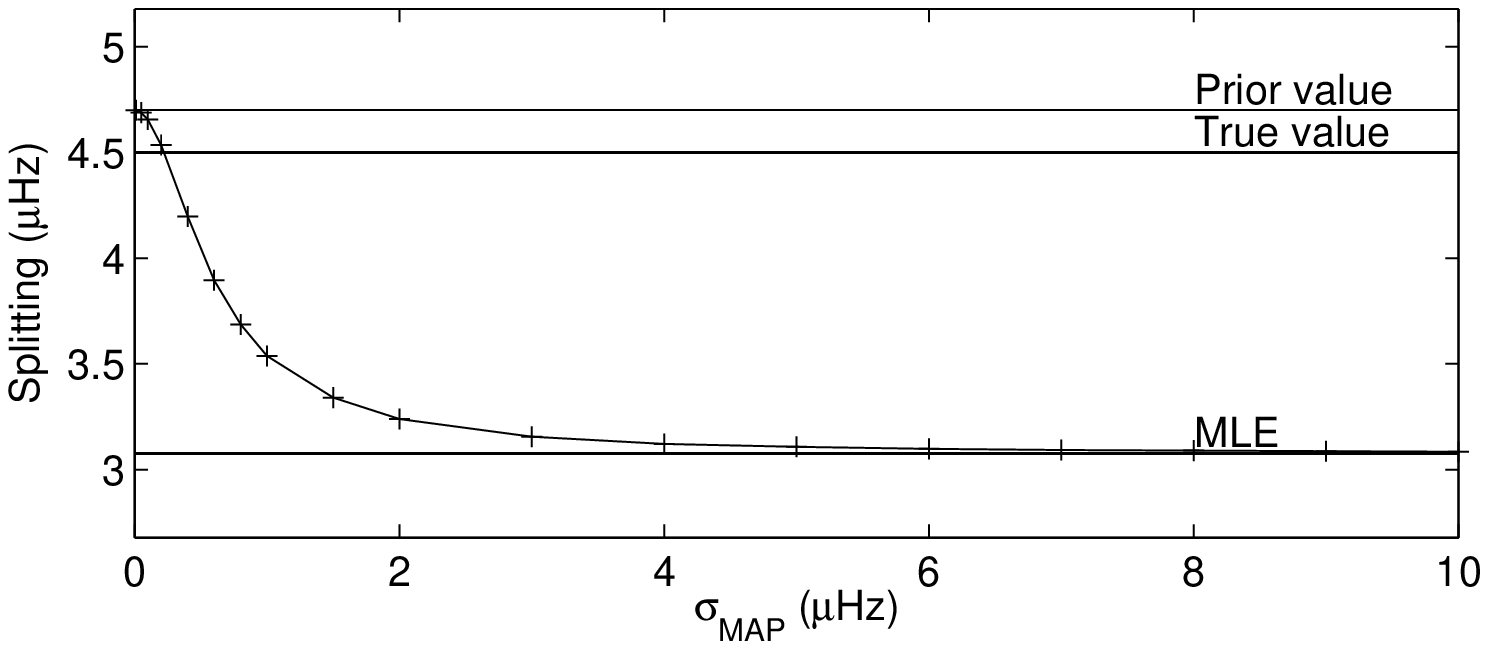}
\caption{Top: excitation (black plain line), simulation of the measured power spectrum with a SNR $= 3$ (gray) and fitting of the power spectrum (black dashed line). Bottom: splitting frequency as function of the prior variance $\sigma\ind{prior}^2$. The output estimate tends to the prior value. }
\label{fig_simu_MAP}
\end{figure}

\subsection{Prudence is required to avoid the traps laid along the way}

In this section, we discuss the importance of carefully selecting the priors. Indeed, as can be anticipated when looking at Eq. \ref{Eq_MAP}, putting a stronger prior (i. e. narrow variance) may introduce a bias: the risk is to obtain exactly the value that we were expecting.    

As an example, Fig. \ref{fig_simu_MAP} (top) presents a simulation of the power spectrum of a single overtone featuring degrees $\ell = [0,2]$, with  SNR$\ =\ 3$, which corresponds to the maximum SNR of HD 181420 (B09). We have tested the MAP fitting procedure by using a set of priors of  the same value but with linearly increasing variance. The only free parameter that we have kept is the splitting frequency. Fig. \ref{fig_simu_MAP}  (bottom) shows the value of the estimated splitting frequency as a function of the prior variance. The ``true'' splitting corresponds to the value used to build the noisy power spectrum, while the ``prior'' splitting is the value that we think to be true and the ``MLE'' splitting is the value obtained with the MLE. When $\sigma\ind{prior}^2$ tends to 0, the output estimate tends to the prior value, whereas as $\sigma\ind{{prior}}^2$ tends to infinity the estimated output tends to the MLE estimate. So, the a priori probability does not help us find theÊ``true'' value when the prior is wrong. It just tends towards the prior value as a function of the prior's strength.

In other terms, the MAP does not push the solution to converge towards the local minimum that we may expect to be present at the ``true'' value of the parameter. Indeed, because of the stochastic nature of the power spectrum and because we have only one data set (the time series is one), the local minimum associated with the ``true'' value does not exist. The ``true'' value is only the mean value obtained on a large set of experiments. 
Thus, i) the use of MAP has to be performed {\it only} with a reliable prior (i.e. solar splitting frequency  $\sim 0.4\ \mu$Hz)
ii) the MAP has to be used only to prevent the fit from converging towards an incorrect solution (i.e. negative splitting).

\section{Application to the global fittings}
\subsection{Putting a prior on the frequency splitting}

The most common problem met with fitting the CoRoT solar-like oscillation spectra is the ``Dirac-like'' convergence (Sect. \ref{sect_struggle}). According to the Bayesian approach, we must use all the information that may be deduced through reliable complementary observations or physical arguments. A proxy of the splitting frequency often can be deduced from the power spectrum itself. Indeed, on the power spectra of the 3 CoRoT targets HD 49933, HD 181420 and HD 181906, an excess power at very low frequency ($< 10\ \mu$Hz) is observed (Fig. 3 in A08 and B09, Fig. 4 in Garcia et al. 2009). This feature is associated with the rotational frequency, which appears directly in the power spectrum thanks to the motion of stellar spots. It corresponds to the star surface rotation. For physical reasons it is clear that the internal stellar rotation does not differ strongly with respect to the surface rotation, otherwise the star would not stay in state of equilibrium (all are main sequence stars). Hence, the splitting prior is determined by this low frequency excess power, with a confidence interval that has to be chosen ad-hoc, as a function of the width of the low-frequency peak (e. g. $\pm 10$\% confidence for Benomar et al. 2009). 

\begin{figure}
\includegraphics[width=8.4cm]{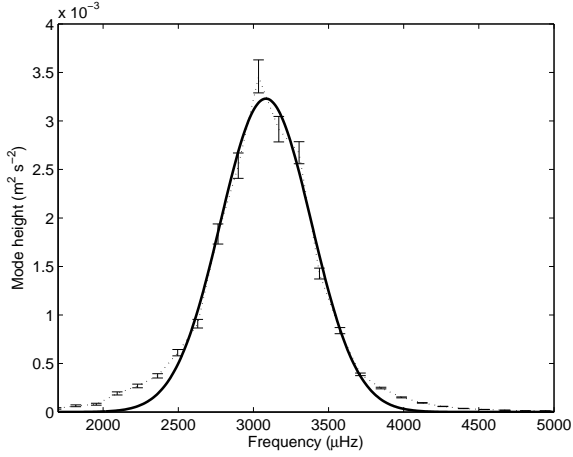}
\caption{Height of the solar modes as a function of the frequency estimated from 1000 days of  GOLF data (black) and its fitting with a Gaussian function (red).}
\label{fig_golf_1}
\end{figure}
\begin{figure}
\includegraphics[width=8.4cm]{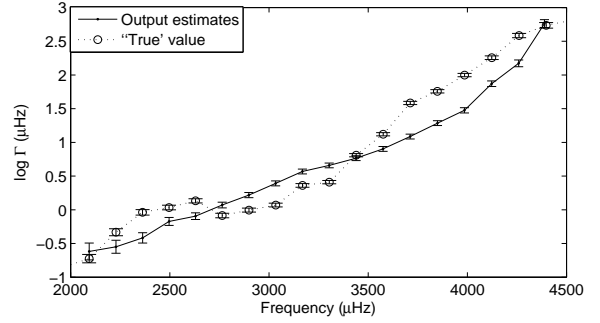}
\caption{Comparison between the mode width values obtained by fitting the modes by groups of 4 overtones (circles and dotted line) on 1000 days of GOLF data and the estimated output values of the same (plain line), obtained with a global fitting by using the Gaussian height hypothesis on a sub-set of 100 days of GOLF data.}
\label{fig_golf_2}
\end{figure}

\begin{figure}
\includegraphics[width=8.4cm]{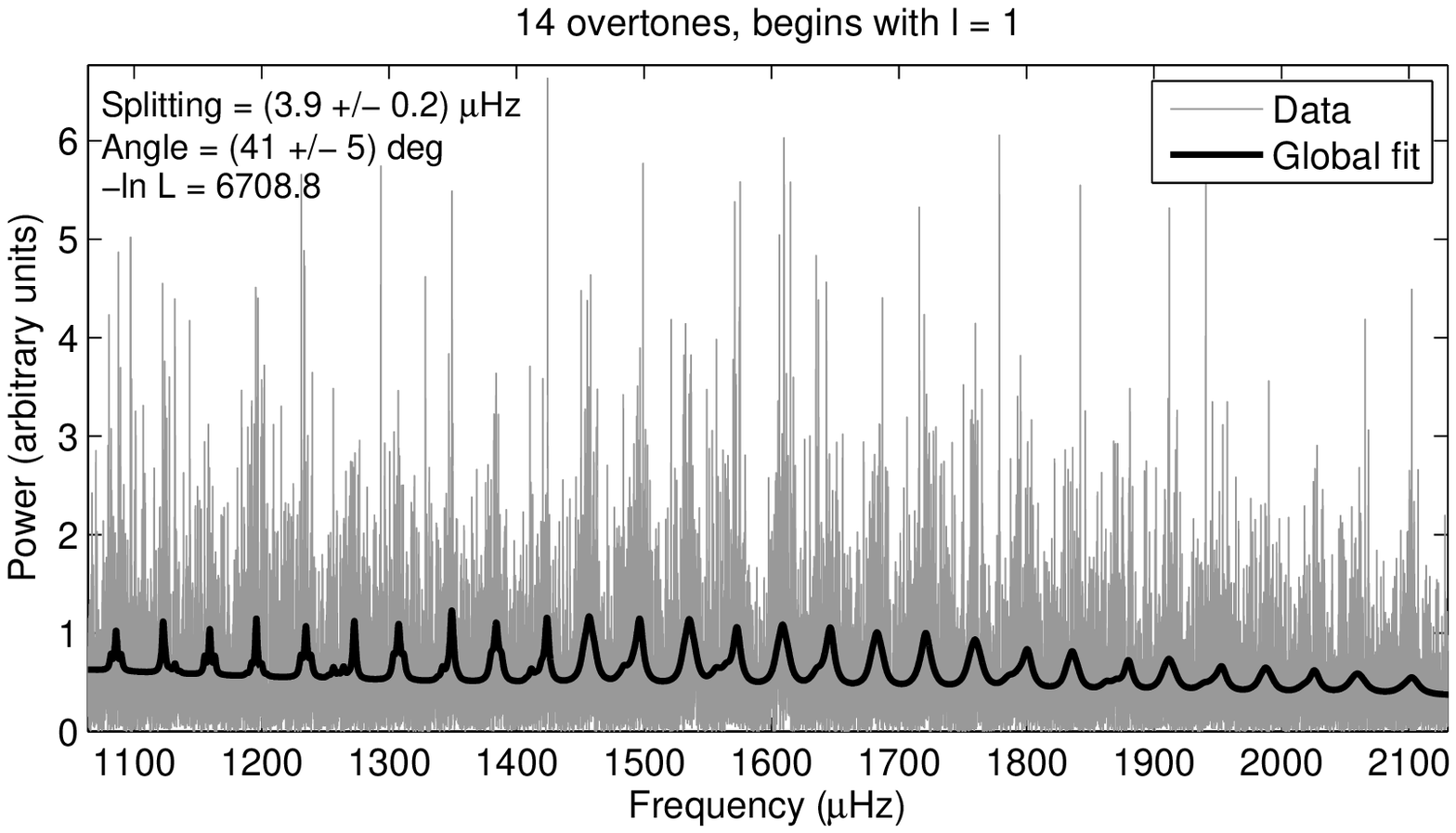}
\includegraphics[width=8.4cm]{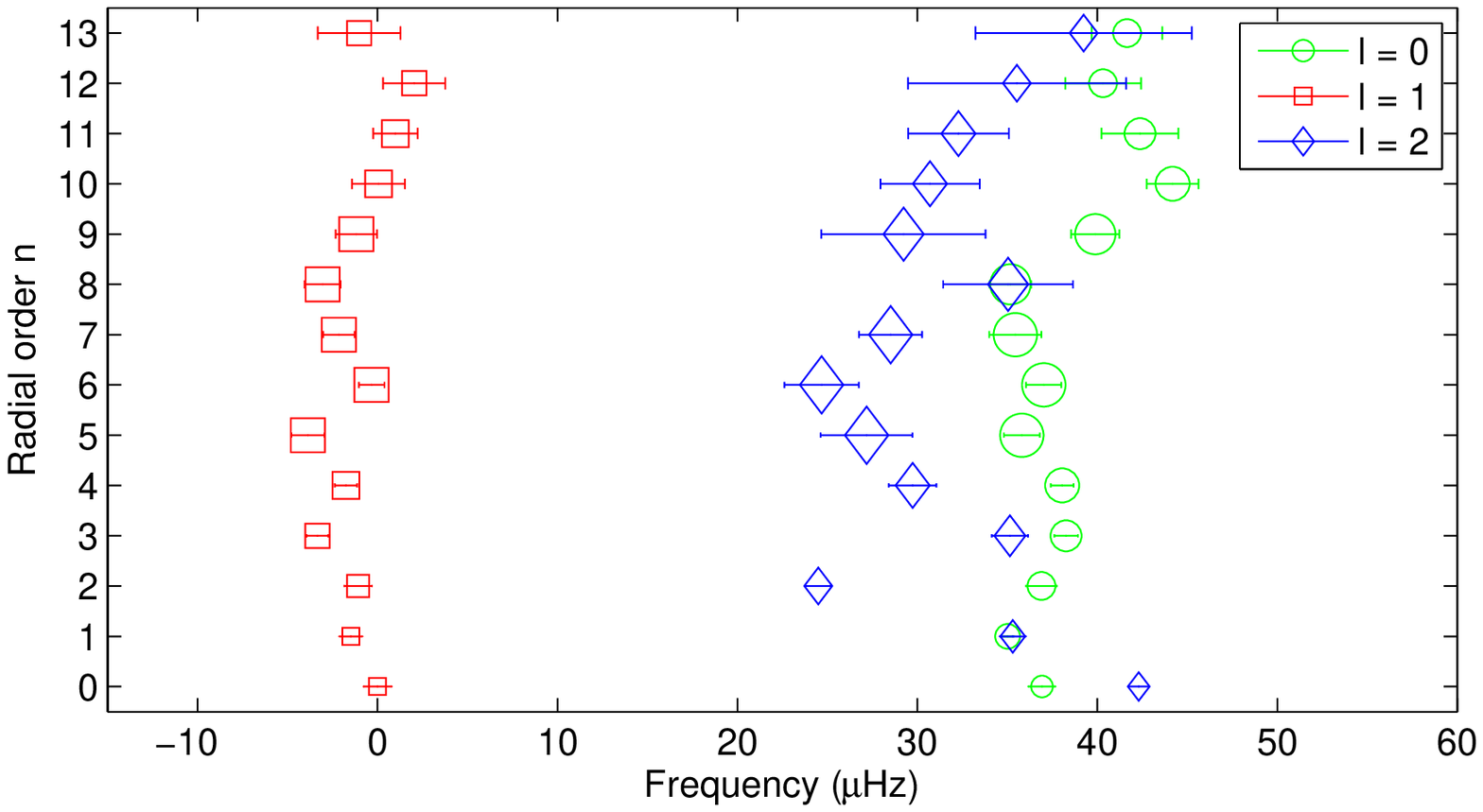}
\includegraphics[width=8.4cm]{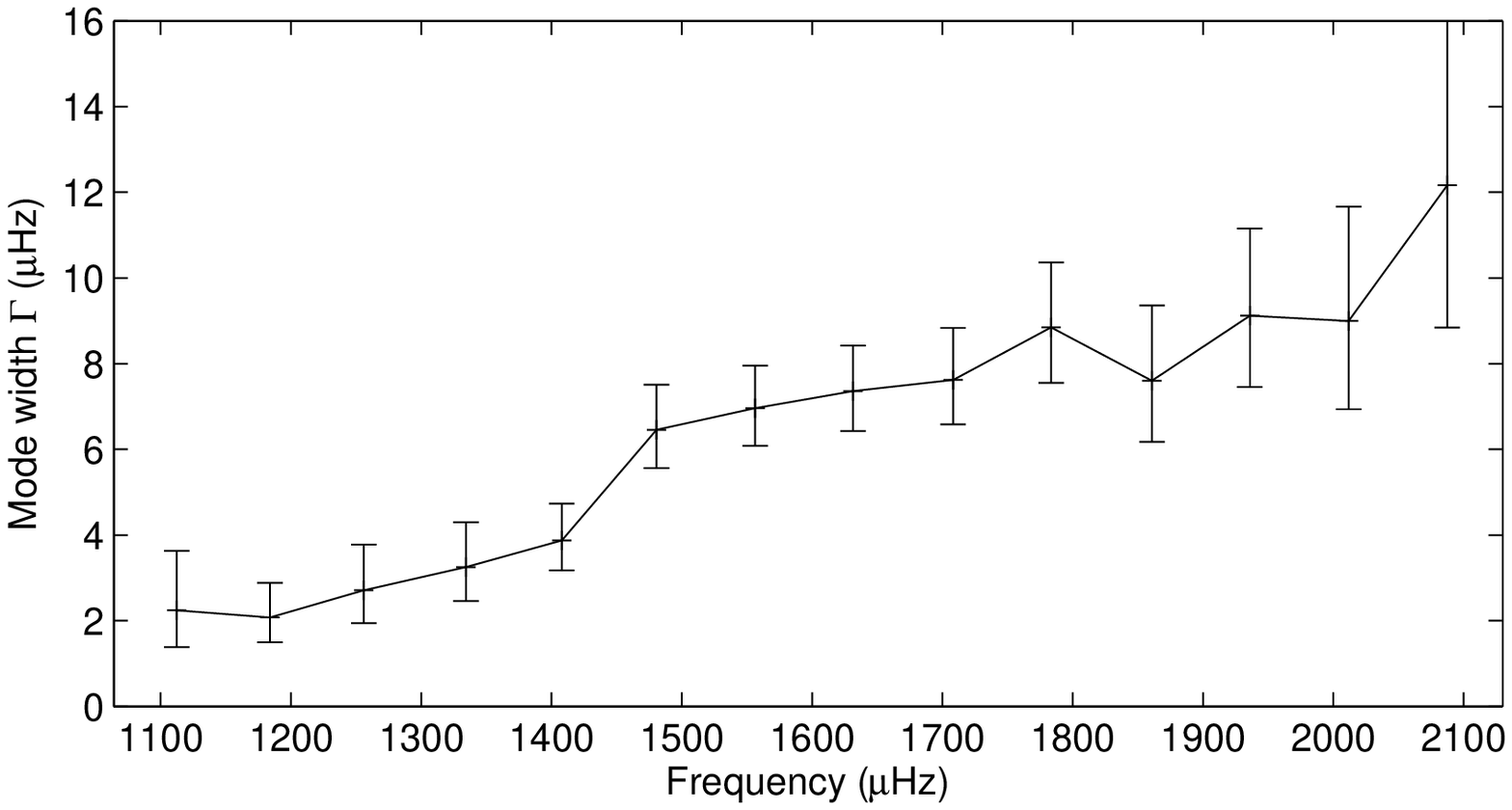}
\caption{Scenario 1. Global fitting of the power spectrum of the solar-like target HD 181420 with the MAP. From top to bottom: 1) power spectrum and fitting, 2) an echelle diagram of the same, 3) estimate of the mode width; its mean slope on a log-log scale is equal to 2.8.}
\label{fig_MAP_S1_181420}
\end{figure}
\begin{figure}
\includegraphics[width=8.4cm]{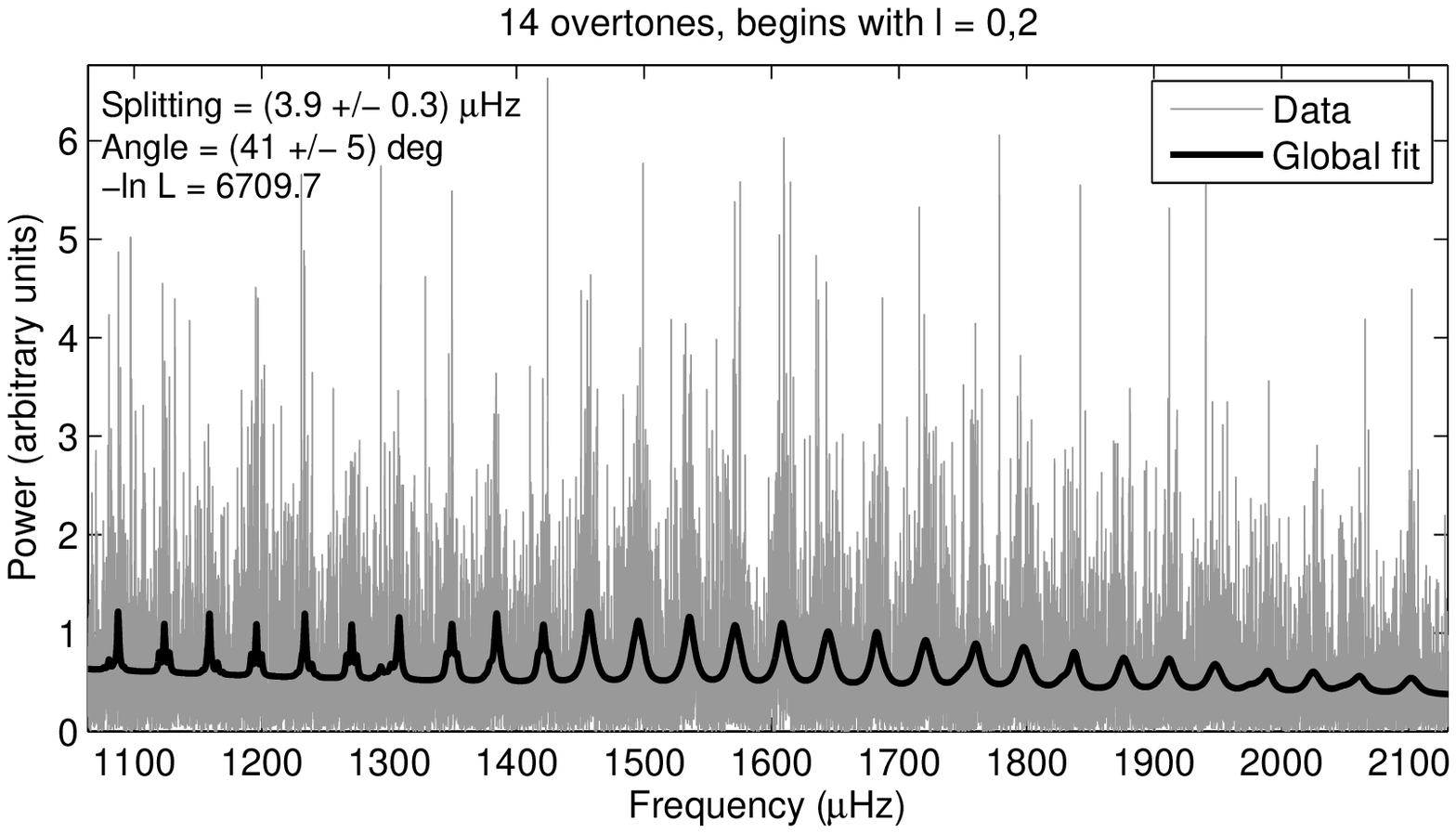}
\includegraphics[width=8.4cm]{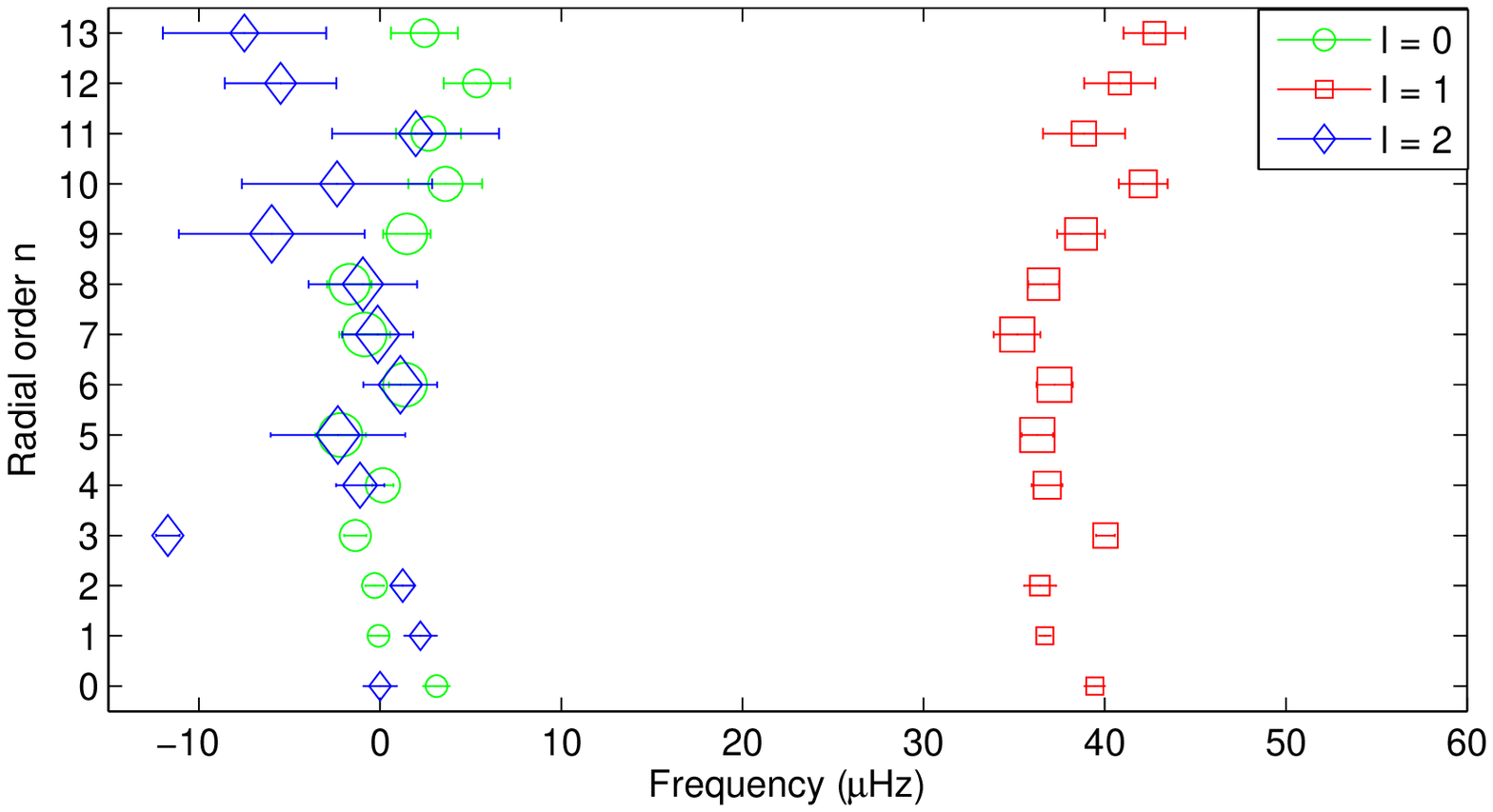}
\includegraphics[width=8.4cm]{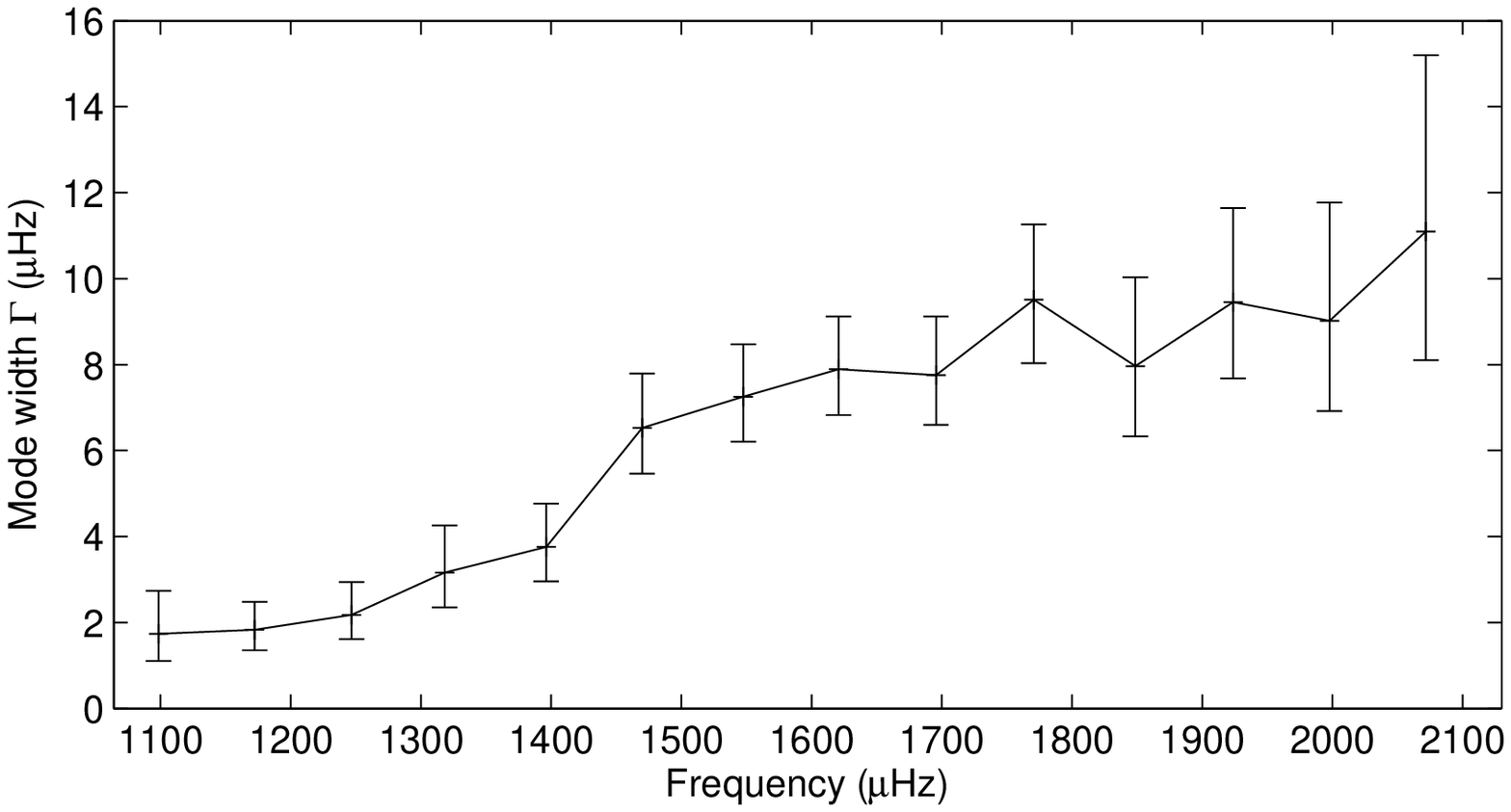}
\caption{Scenario 2. Global fitting of the power spectrum of the solar-like target HD 181420 with the MAP. From top to bottom: 1) power spectrum and fitting, 2) an echelle diagram of the same, 3) estimate of the mode width; its mean slope on a log-log scale is equal to 3.1.}
\label{fig_MAP_S2_181420}
\end{figure}

\subsection{Putting a constraint on the height}

The second prior addresses the original purpose of the paper. We started by considering that the mode amplitude varies as a continuous function of the frequency. This is justified by theoretical studies (\citealt{Houdek_99}, \citealt{Samadi_01} and \citealt{Samadi_07}) and has been observed in solar seismological data (e.g. \citealt{Chaplin_98}). 
The mode amplitude of a single p-mode is defined by its integral $A=\sqrt{\pi \Gamma H}$. As shown in \citet{Toutain_Appourchaux_94}, the mode height and width determinations are strictly correlated. So, constraining one of them is sufficient to constrain the amplitude. 

In solar seismological data (e.g. GOLF), the height presents a ``bell'' shaped dependence as a function of the frequency (Fig. \ref{fig_golf_1}), while the width presents a more complex behavior: 2 slopes and a plateau (Fig. \ref{fig_golf_2}). On CoRoT data (e.g. HD 49933 in A08) the smoothed power spectrum shows that the height mainly follows a bell shaped profile, as for the Sun. It reflects the fact that the oscillation spectrum presents a beginning, an ending and a maximum in between. On the other hand, contrarily to the GOLF solar data in which the mode width is measurable with the naked eye, in the CoRoT data it is impossible to give an estimate of the width trend, because of the SNR.    

It appeared logical to take such information into account. The difficulty is to find a simple mathematical and computational translation of it. The first idea is to fix the height close to a bell shaped profile. The drawback of such an approach is to introduce several hyper parameters to describe the bell shaped profile in the fitting process, which goes against the aim of reducing both the total number of parameters and the computational time. Therefore, we propose to replace the one height per overtone estimate by a continuous analytical function of the frequency, whose parameters are determined in the fitting process.    

Figure \ref{fig_golf_1} presents the solar mode heights as measured in 3 years of GOLF data, corresponding to a period of solar minimum activity. We have estimated the mode parameters by fitting the power spectrum with the MLE in groups of 4 overtones: 27 modes were fitted in the frequency range $[1650,5000]\ \mu$Hz. The purpose of the plot is to show that the height is well approximated by a simple Gaussian profile in the central part of the oscillation spectrum, that is, in the frequency range $[2500,3800]\ \mu$Hz, which corresponds to 10 overtones. A better fit would be obtained by complicating the fitting function, such as a combination of a Gaussian and an Lorentzian profile (pseudo-Voigt), but it is useless for the following analysis. As a comparison, in the HD 181420 CoRoT case, the frequency range in which the modes are high enough to be fitted is $[1060, 2250]\ \mu$Hz, i.e. 14 overtones. So, by assuming the solar-like star to behave like the Sun, the ``Gaussian height approximation'' (hereafter GHA) is reliable for most of the power spectrum.    

To test the validity of this approximation, a global fit taking into account 18 solar overtones was performed on 90 days of GOLF data in the frequency range $[2000,4500]\ \mu$Hz, which is larger that the confidence interval of the Gaussian fit. The comparison between the quadruplet mode fitting on the 1000 days of data and the global fit with the 90 days of data is done in Fig. \ref{fig_golf_2}. The width estimate is different in the details, but the global trend is conserved. The mean slope of the bandwidth on a log-log scale is equal to 3.84 with the not-approximated approach, whereas it is equal to 3.99 with the GHA, which represent an incorrect isestimate of almost 4\%. Note that such an operation is a MAP approach with $\sigma\ind{prior}^2 = 0$; the probability that height follows a Gaussian profile is equal to 1. 

The possibility of putting priors on the frequency localization was not considered, even though it could be done reliably by using the position of the power maxima measured on a smoothed power spectrum. The reason is that it has been observed that the mode frequencies are parameters for which no problems of convergence are met.
Moreover, such a constraint on the frequency position implies a prior on the small separations $[0,1]$ and $[0,2]$, for which we do not have valid constraint: until now, in all CoRoT solar-like observations, it was impossible to determine which of the peaks corresponds to degree $\ell = 0,2$ or $\ell=1$.  

\begin{figure}
\includegraphics[width=8.4cm]{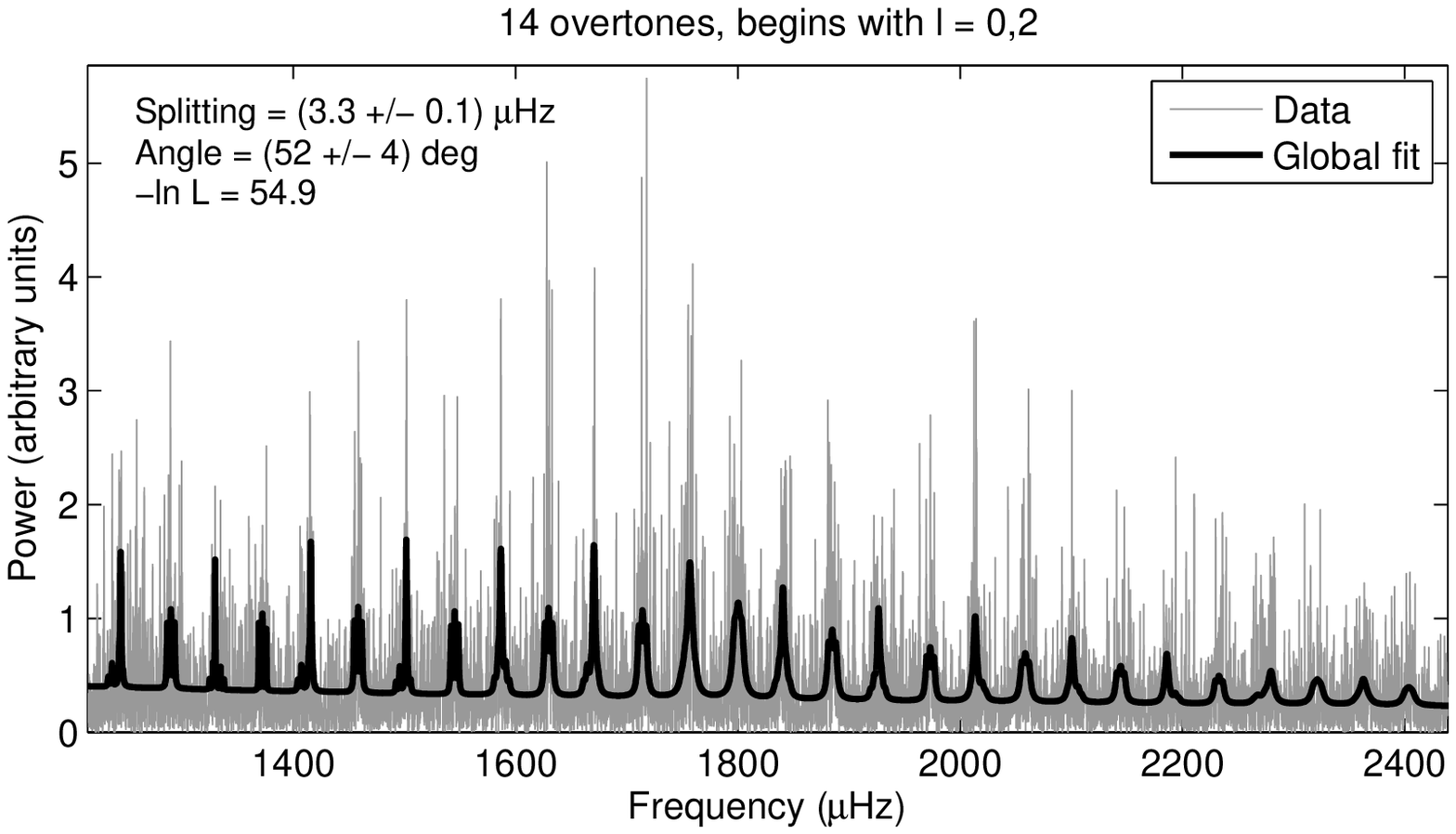}
\includegraphics[width=8.4cm]{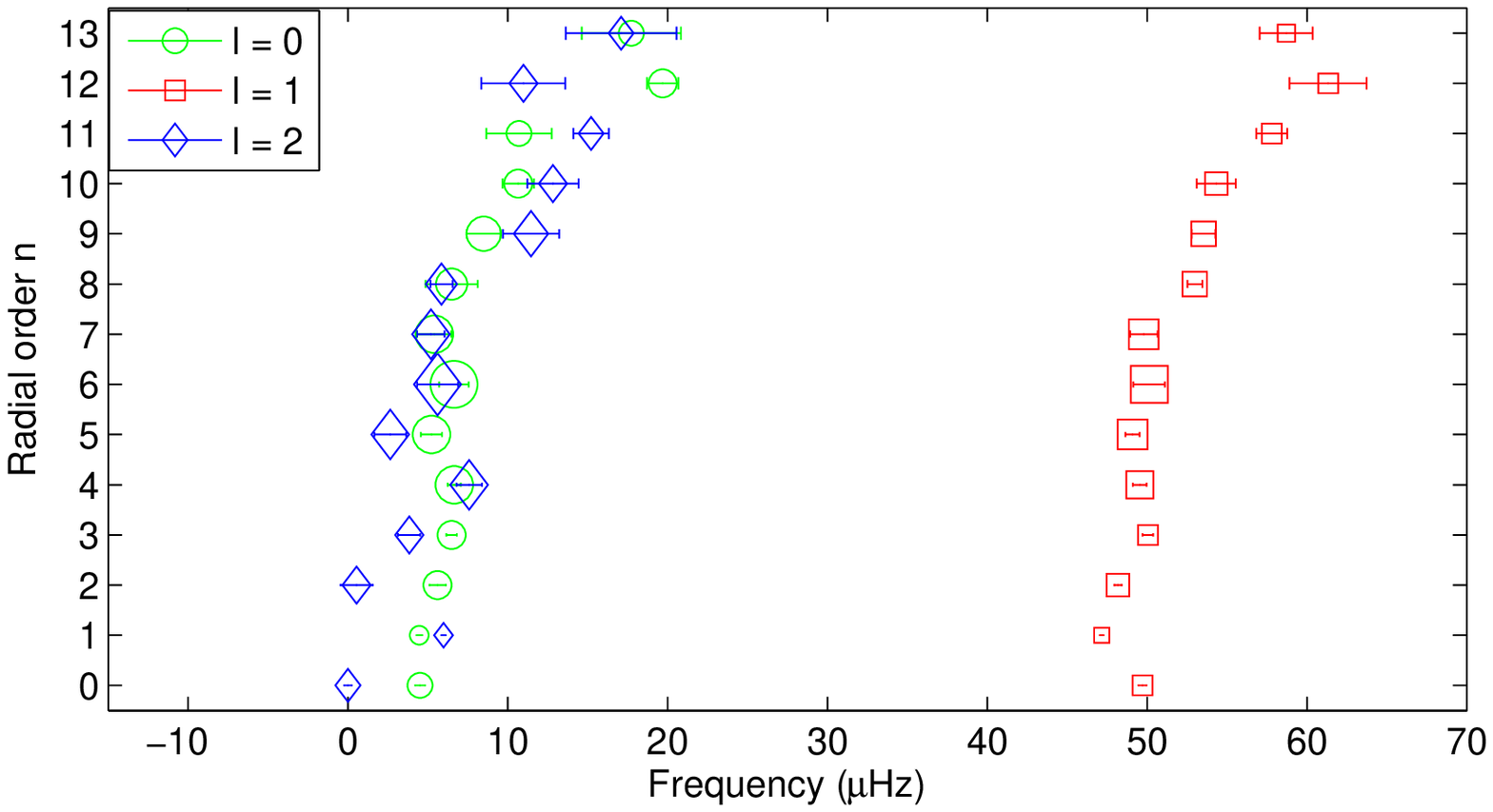}
\includegraphics[width=8.4cm]{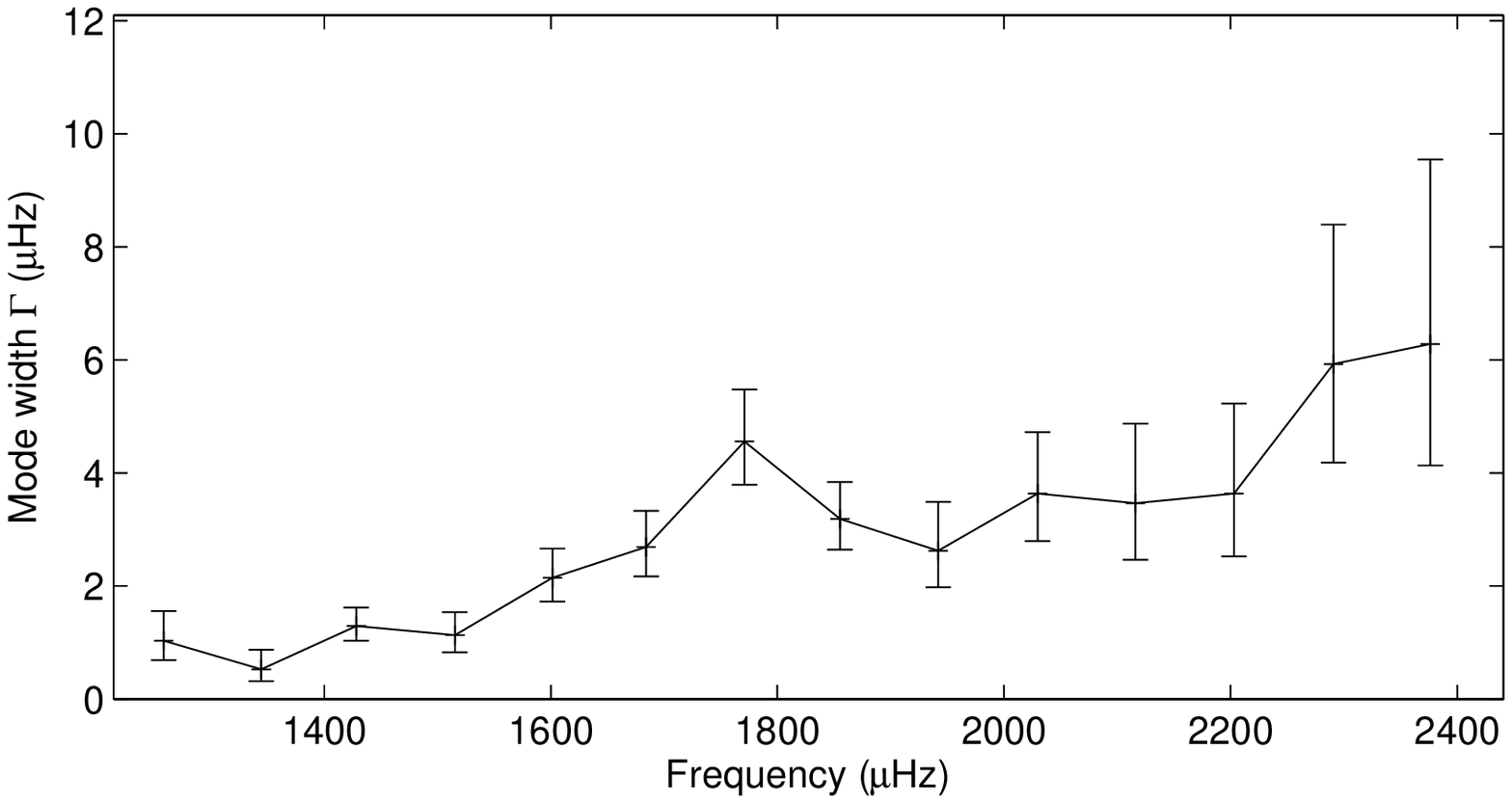}
\caption{Scenario 1. Global fitting of the power spectrum of the solar-like target HD~49933 with the MAP. From top to bottom: 1) power spectrum and fitting, 2) an echelle diagram of the same, 3) estimate of the modes width; its mean slope in log-log scale is equal to 3.2 on a log-log scale.}
\label{fig_MAP_S1_49933}
\end{figure}
\begin{figure}
\includegraphics[width=8.4cm]{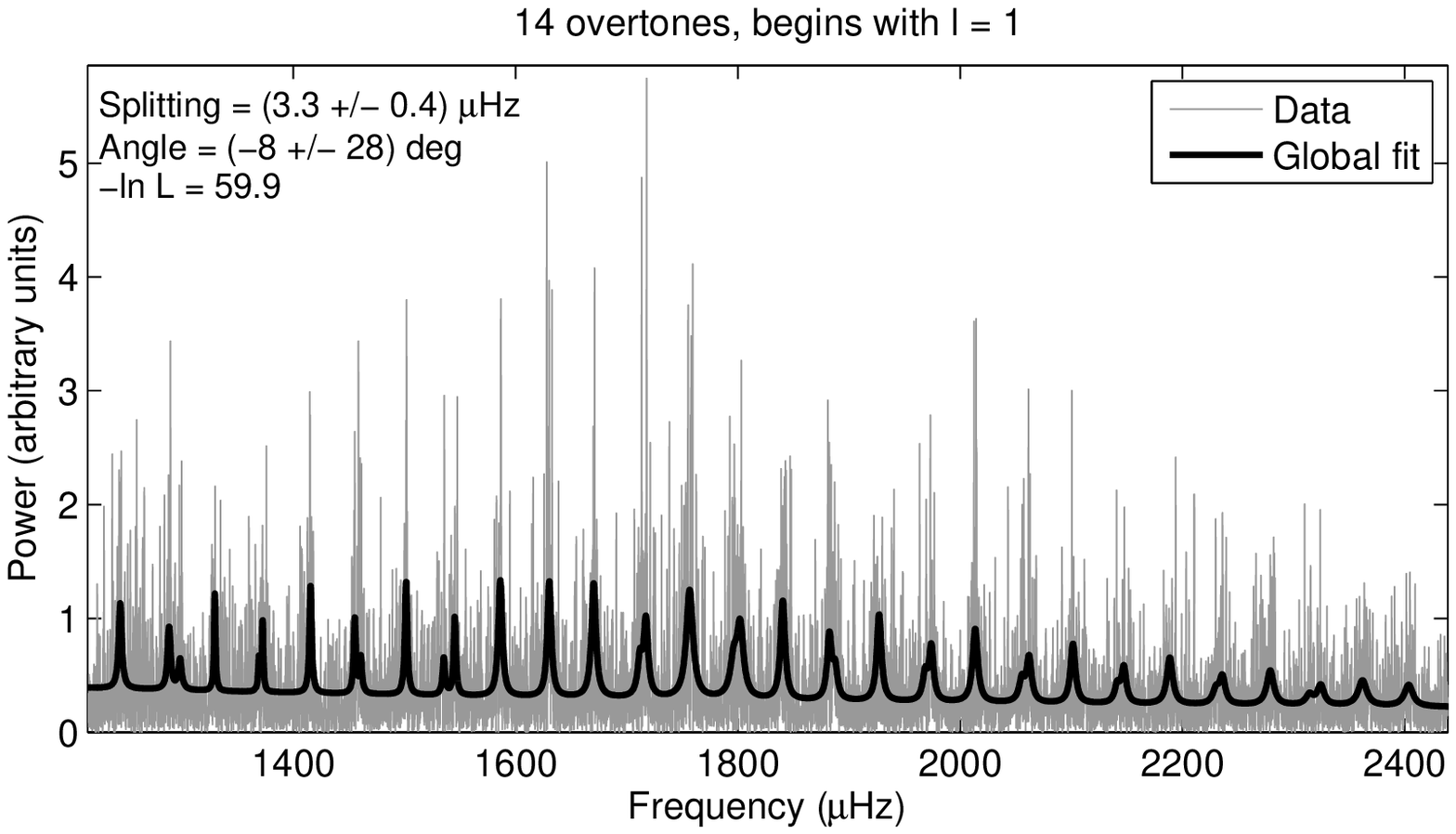}
\includegraphics[width=8.4cm]{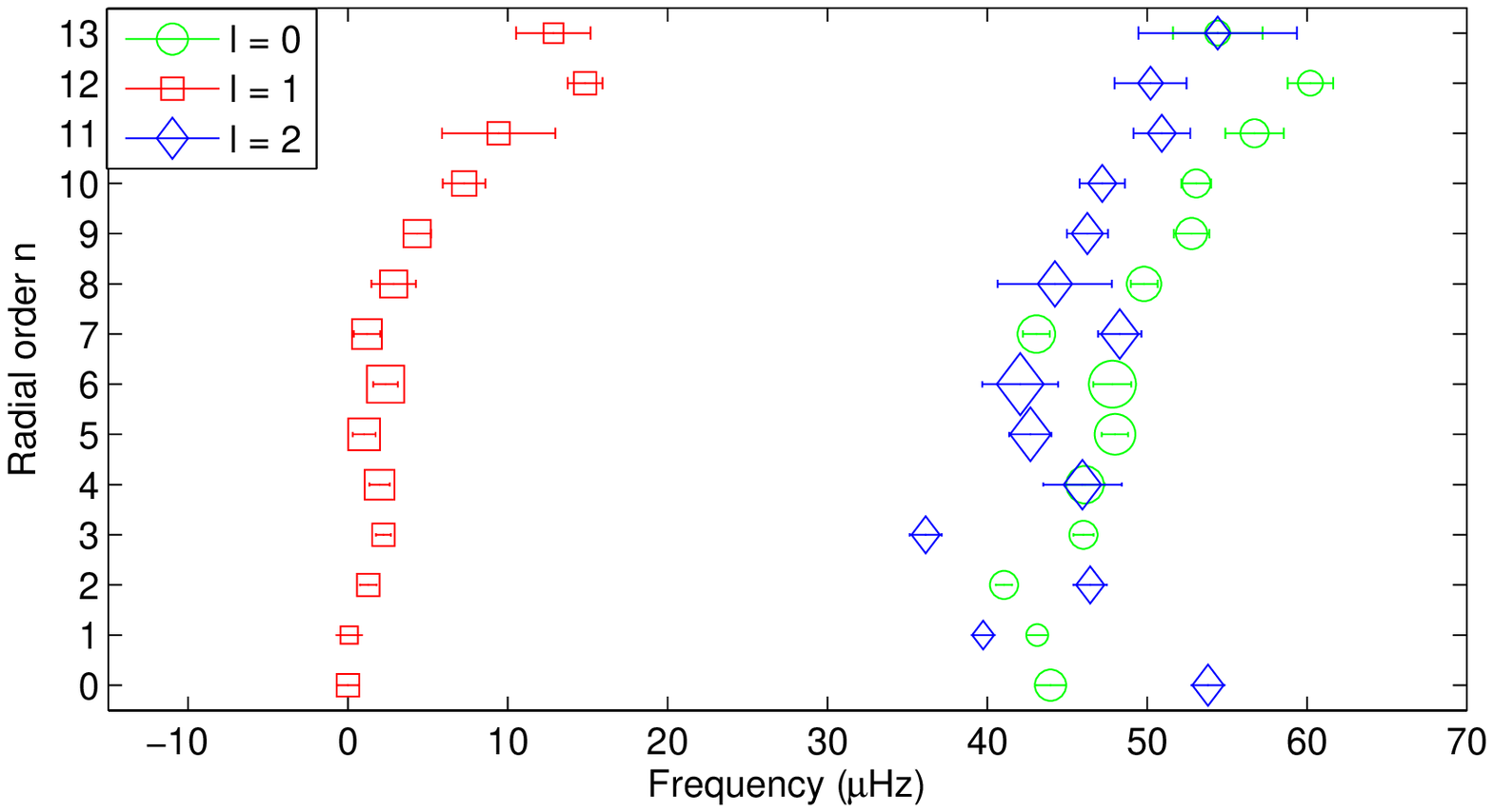}
\includegraphics[width=8.4cm]{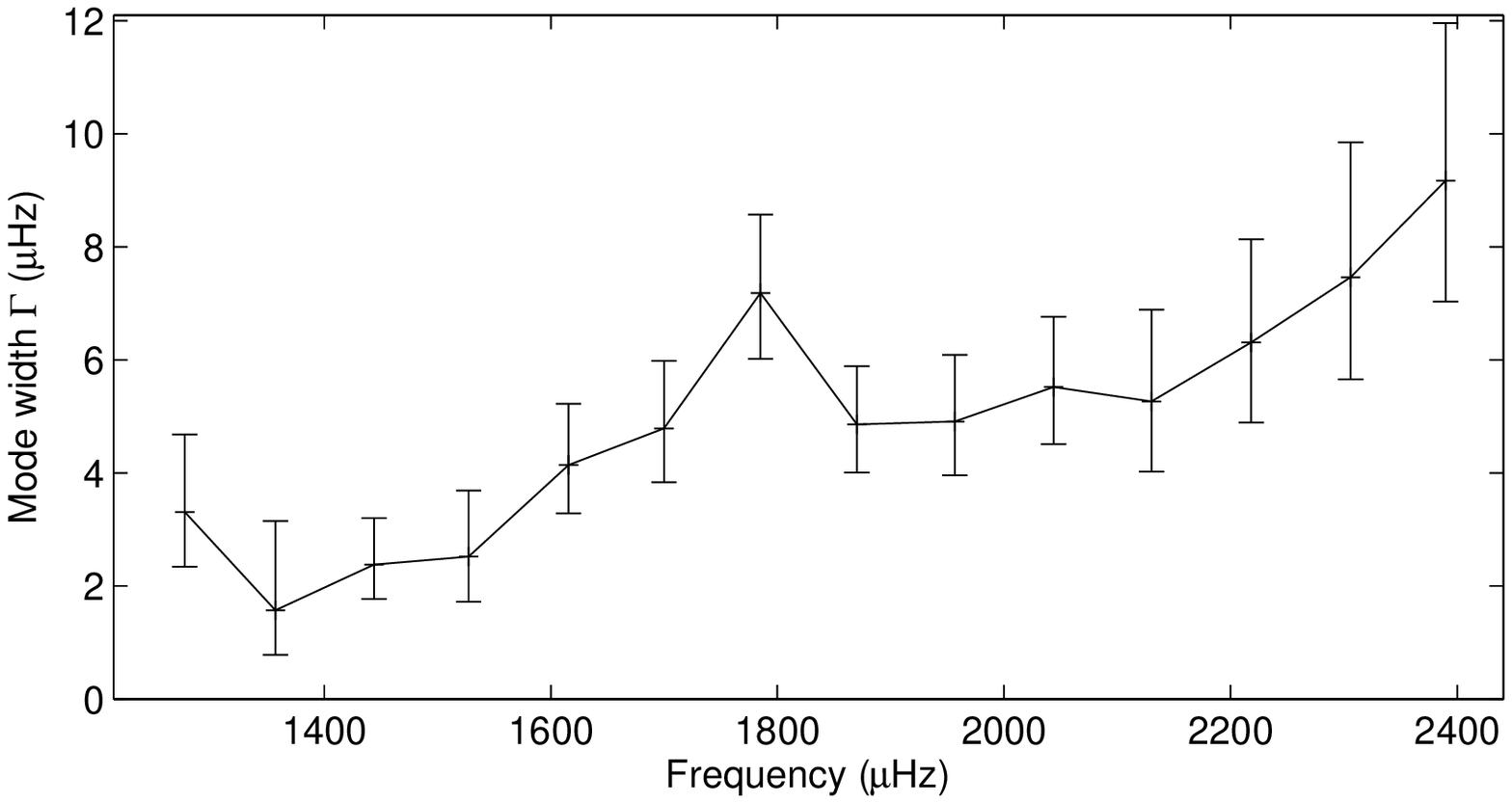}
\caption{Scenario 2. Global fitting of the power spectrum of the solar-like target HD~49933 with the MAP. From top to bottom: 1) power spectrum and fitting, 2) an echelle diagram of the same, 3) estimate of the mode width; its mean slope in log-log scale is equal to 2.5 on a log-log scale.}
\label{fig_MAP_S2_49933}
\end{figure}

\section{Application to CoRoT targets HD~181420 and HD~49933}

We applied such a method to the global fitting of the two CoRoT solar-like targets that were globally fitted with MLE: HD~181420 (cf B09) and HD~49933 (cf A08). In this section, we present the specific choices of the priors, then the astrophysical results.   

\subsection{Priors on the splitting frequency, the Harvey profile and the height}

According to the excess power observed in both stars at low frequency, the splitting prior was set at the maximum value of the peaks. For HD~181420, a quite large excess power appears in the frequency range $[3.2, 6]\ \mu$Hz (cf B09), with a maximum at 4.2 $\mu$Hz; the prior splitting frequency was fixed at $4.2 \pm 0.5\ \mu$Hz. For HD~49933, the excess power appears as a peak at almost $3.3\ \mu$Hz (cf A08);  the prior splitting frequency was fixed at $3.3 \pm 0.4\ \mu$Hz.  

Secondly, the low frequency stellar noise is well fitted by a sum of 3 background components, i.e. 3 Lorentzian functions of the frequency, centered on 0 and with 3 different amplitudes and widths \citep{Harvey_85}. In our case, we do not consider the too low frequency range ($\nu < 600\ \mu$Hz), hence one background component is sufficient. The power term $p$ (see Eq. \ref{Eq_harvey}) is set to 2. Despite the simple analytic expression, the background is the most sensitive contribution of the global fitting to be determined. Indeed, the consequences of an incorrect fitting are damaging for the oscillation parameter estimates. As an example, overestimating the background leads to an underestimate of the mode amplitudes and may strongly change the output values for the height even more for the width. Therefore, to diminish the risk of incorrectly estimating the background, we fitted it alone over a wide frequency range $[600,5000]\ \mu$Hz, where the oscillation mode energy is negligible and the interval large enough to constrain the background parameters. Then, when fitting the oscillation spectrum, these parameters are set as free with the previous estimates as priors. The tolerance is set to 5\% of the prior values, to permit small fluctuations due to the influence of the oscillation power when fitting the background profile alone. 

For the GHA, the prior value of the width of the Gaussian function $\sigma\ind{H}^2$ is set to the approximate width at half maximum estimated on a smoothed power spectrum over 25 $\mu$Hz. For HD~181420 and HD~49933 it was set as $\sigma\ind{H}=(425 \pm 200)\ \mu$Hz. Moreover, a constraint was placed on the centering of the Gaussian function on the frequency axis. In the same way, it arises from the position of a smoothed power spectrum (over 350 $\mu$Hz). For HD~181420 it was set at $\nu\ind{H} =(1535 \pm 200)\ \mu$Hz, while for HD~49933 it was set at $\nu\ind{H} = (1760 \pm 200)\ \mu$Hz.

\subsection{Results: the mode width increases with frequency}
\label{results}
For each star, two mode identifications are tested because of our inability to identify with the naked eye the odd from the even modes. For HD~181420, the scenario deals with the identification of the peak at almost 1500 $\mu$Hz: it corresponds to the pair $\ell = (0,2)$ in scenario 1 and to $\ell = 1$ in scenario 2. For HD~49933, the scenario deals with the identification of the first considered peak, at almost 1250 $\mu$Hz: it corresponds to the pair $\ell = (0,2)$ in scenario 1 and to $\ell = 1$ in scenario 2. Figs. \ref{fig_MAP_S1_181420},  \ref{fig_MAP_S2_181420},  \ref{fig_MAP_S1_49933} and  \ref{fig_MAP_S2_49933} present the results of the global fits obtained with the GHA for HD~181420 and HD~49933.

\begin{figure}
\includegraphics[width=8.4cm]{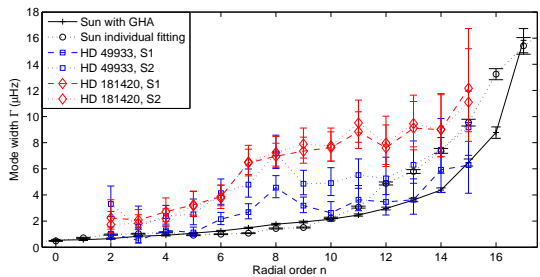}
\caption{Comparison between the mode widths as estimated on the CoRoT solar-like targets HD~49933 (blue squares) HD~181420 (red circles) and on the Sun (black plus). The dashed lines indicate scenario 1, while the dotted lines scenario 2. For the Sun, the plain line results from the GHA while the dashed line from the ``quadruplet'' fitting (Sect. 4.2).}
\label{fig_width}
\end{figure}

Beyond the fact that the Dirac-like convergence logically disappeared, 
the main result is the width trend as function of the frequency: in the 2 stars, whatever the scenario is, the mode width increases continuously. For HD~181420 the mean slope of the width on a log-log scale is equal to 2.8 in scenario 1 and 3.1 in scenario 2, while for HD~49933 the mean slope is equal to 3.2 and 2.5. These values are lower than the GOLF solar value, where the mean slope was estimated to be 4 in the central part of the oscillation spectrum. 
However, in Fig.~\ref{fig_width} the widths of the two stars and the Sun are represented as functions of the radial order $n$, instead of the frequency, to take into account the scaling factor due to the different size and density of the 3 stars. With such a representation, the mode width frequency dependences look more similar. Note that the slight excess widths observed at the maximum power of the 2 star spectra (radial order $n=8$) are the probable signature of the fact that the height profile differs slightly from a perfect Gaussian trend. It is probably sharper in this frequency range. 

In addition, note that the mode frequency estimates changed noticeably between the MLE and the MAP fits, as illustrated for scenario 1 of HD~181420 (Fig. \ref{fig_MLE_181420}). In the MAP approach, modes with degrees $\ell=0$ and $\ell=2$ are clearly disconnected (mean small separation of about 5-10 $\mu$Hz) while with the MLE the two modes interlace. Also, a careful analysis of the small separation $0-2$, particularly in scenario 1 of HD~181420 and scenario 2 of HD~49933, shows that modes $\ell=0$ and $\ell=2$ are often inverted. A careful prior should be inserted in the fitting process in order to fix the small separation in both cases ($\ell=0$ at left of $\ell=2$ and {\it vice versa}); then the likelihood value would determine which solution is the most likely. Such an analysis has to be carried out very carefully and the MCMC approach would be preffered to more accurately explore the parameter space. This is beyond the scope of the present work.
 
\section{Conclusion}
According to the suggestions of Appourchaux (2008) we have applied the maximum a posteriori approach to the global fitting of solar-like power spectra. It consists of a regularized maximum likelihood estimator, and hence is easy to implement. In particular, we have hypothesized that the mode height follows a Gaussian profile as function of the frequency. This was justified by the inability of the MLE to correctly fit the whole oscillation spectrum (Fig. 1). In other words, it is reasonable to differently process data with SNR$Ê=Ê3-10$ than solar data with SNRÊ$ >Ê100$.

This method allowed us to globally fit the power spectra of 2 solar-like targets HD~181420 and HD~49933 over 14 overtones. The main result of such a method is that the mode width increases with frequency, as for the Sun. Moreover, it appears that the mean slope of the width as a function of the frequency on a log-log scale is similar to the solar value.

About the method, we conclude that with a power spectrum where the SNR lies between 3 and 10, a Bayesian approach is unavoidable, unless we are only able to properly fit the central part of the spectrum. 
The MAP estimator is very easy to apply, but as highlighted in Sect. 3.2, it has to be handled carefully. Such a method is a powerful tool to quickly and efficiently fit large samples such as those from Kepler (NASA) or for PLATO (ESA Cosmic Vision program).

\section{Acknowledgements}
We thank the CNES for the short term contract which allowed P. Gaulme to carry out the present work. The authors also thank F.X. Schmider and A. Ferrari (Fizeau, OCA, Nice) for discussions about the global fitting of oscillation spectra in the solar case and J. Leibacher for the patience in carefully reading the present manuscript.

\begin{table*}
\centering
 \caption{Output parameters from the global fitting of HD~181420.}
\begin{tabular}{| c c | c c|c c|}
\hline
\hline
HD~181420 & & Scenario 1  &&Scenario 2&\\
\hline
Degree & Order& Frequency & Uncertainty&Frequency & Uncertainty \\
& & $\mu$Hz    &$\mu$Hz &  $\mu$Hz    &$\mu$Hz\\
 \hline
0 &  0 & 1122.8 & 0.8 & 1087.4 & 0.7 \\
0 &  1 & 1195.9 & 0.4 & 1159.2 & 0.6 \\
0 &  2 & 1272.7 & 0.9 & 1234.0 & 0.5 \\
0 &  3 & 1349.1 & 0.7 & 1307.9 & 0.6 \\
0 &  4 & 1423.9 & 0.6 & 1384.4 & 0.6 \\
0 &  5 & 1496.7 & 1.0 & 1457.1 & 1.4 \\
0 &  6 & 1572.9 & 1.0 & 1535.6 & 0.9 \\
0 &  7 & 1646.3 & 1.4 & 1608.4 & 1.4 \\
0 &  8 & 1721.0 & 1.2 & 1682.6 & 1.2 \\
0 &  9 & 1800.7 & 1.3 & 1760.7 & 1.3 \\
0 & 10 & 1880.0 & 1.4 & 1837.9 & 2.0 \\
0 & 11 & 1953.2 & 2.1 & 1911.9 & 1.8 \\
0 & 12 & 2026.2 & 2.1 & 1989.6 & 1.8 \\
0 & 13 & 2102.5 & 2.0 & 2061.7 & 1.9 \\
\hline
1 &  0 & 1085.9 & 0.8 & 1123.7 & 0.6 \\
1 &  1 & 1159.4 & 0.6 & 1196.0 & 0.3 \\
1 &  2 & 1234.8 & 0.8 & 1270.7 & 0.9 \\
1 &  3 & 1307.5 & 0.6 & 1349.3 & 0.5 \\
1 &  4 & 1384.1 & 0.6 & 1421.1 & 0.8 \\
1 &  5 & 1457.0 & 0.9 & 1495.5 & 0.9 \\
1 &  6 & 1535.5 & 0.7 & 1571.5 & 1.0 \\
1 &  7 & 1608.7 & 0.9 & 1644.4 & 1.3 \\
1 &  8 & 1682.8 & 1.0 & 1720.9 & 0.9 \\
1 &  9 & 1759.7 & 1.1 & 1797.9 & 1.3 \\
1 & 10 & 1835.9 & 1.5 & 1876.4 & 1.3 \\
1 & 11 & 1911.8 & 1.2 & 1948.1 & 2.3 \\
1 & 12 & 1987.9 & 1.7 & 2025.1 & 2.0 \\
1 & 13 & 2059.8 & 2.3 & 2102.0 & 1.7 \\
\hline
2 &  0 & 1128.2 & 0.6 & 1084.3 & 0.9 \\
2 &  1 & 1196.1 & 0.7 & 1161.5 & 0.9 \\
2 &  2 & 1260.3 & 0.7 & 1235.5 & 0.6 \\
2 &  3 & 1346.0 & 1.0 & 1297.5 & 0.6 \\
2 &  4 & 1415.6 & 1.3 & 1383.2 & 1.3 \\
2 &  5 & 1488.0 & 2.6 & 1456.9 & 3.7 \\
2 &  6 & 1560.5 & 2.1 & 1535.4 & 2.0 \\
2 &  7 & 1639.4 & 1.7 & 1609.1 & 2.0 \\
2 &  8 & 1720.9 & 3.6 & 1683.3 & 3.0 \\
2 &  9 & 1790.1 & 4.6 & 1753.3 & 5.1 \\
2 & 10 & 1866.6 & 2.8 & 1831.9 & 5.3 \\
2 & 11 & 1943.1 & 2.8 & 1911.2 & 4.6 \\
2 & 12 & 2021.4 & 6.1 & 1978.8 & 3.1 \\
2 & 13 & 2100.1 & 6.0 & 2051.8 & 4.5 \\
\hline
\hline
Order& &Width & Uncertainty&Width &Uncertainty \\
&& $\mu$Hz   &$\mu$Hz &  $\mu$Hz  &  $\mu$Hz  \\
 \hline
    0 &  &         2.24 & +1.39/-0.86 &         1.74 & +1.00/-0.63 \\
    1 &  &         2.08 & +0.81/-0.58 &         1.83 & +0.65/-0.48 \\
    2 &  &         2.71 & +1.06/-0.76 &         2.18 & +0.76/-0.56 \\
    3 &  &         3.25 & +1.05/-0.79 &         3.17 & +1.10/-0.81 \\
    4 &  &         3.88 & +0.86/-0.70 &         3.76 & +1.01/-0.80 \\
    5 &  &         6.46 & +1.05/-0.90 &         6.53 & +1.27/-1.06 \\
    6 &  &         6.96 & +1.00/-0.87 &         7.25 & +1.22/-1.04 \\
    7 &  &         7.36 & +1.07/-0.93 &         7.89 & +1.23/-1.06 \\
    8 &  &         7.62 & +1.21/-1.04 &         7.76 & +1.36/-1.16 \\
    9 &  &         8.85 & +1.52/-1.30 &         9.51 & +1.75/-1.48 \\
   10 &  &         7.60 & +1.76/-1.43 &         7.97 & +2.06/-1.64 \\
   11 &  &         9.12 & +2.04/-1.67 &         9.46 & +2.18/-1.77 \\
   12 &  &         9.00 & +2.67/-2.06 &         9.02 & +2.75/-2.11 \\
   13 &  &        12.17 & +4.58/-3.33 &        11.10 & +4.10/-2.99 \\
   \hline
\hline
Height&$H\ind{H}$ (ppm$^2\mu$Hz$^{-1}$) &0.62 & +0.08/-0.07 &   0.62 & +0.10/-0.09\\
&$\nu\ind{H}$ ($\mu$Hz) &1403.5 & 77.7 & 1324.2 & 92.0\\
&$\sigma\ind{H}$ ($\mu$Hz) &559.7 & 72.3 &  620.4 & 77.6\\
\hline
Background&B  (ppm$^2\mu$Hz$^{-1}$)&0.28 & +0.01/-0.01 &         0.28 & +0.02/-0.02 \\
&K (ppm$^2\mu$Hz$^{-1}$)&82.8 & +3.8/-3.6 &        82.6 & +7.4/-6.8 \\
&C ($10^{-6}$ s$^2$)&208.6 & +9.6/-9.1 &       209.0 & +18.8/-17.2 \\
\hline
Splitting&$\nu\ind{S}$ ($\mu$Hz) &4.0 &  0.2 &   3.9 &  0.3\\
Inclination &$i$ (deg)&41 &    5 &     41 &    5\\
Likelihood&$l\ind{MAP}$&6708.8 & & 6709.7 &\\

\hline
\end{tabular}
\label{tab:campagne}
\end{table*}

\begin{table*}
\centering
 \caption{Output parameters from the global fitting of HD~49933.}
\begin{tabular}{|c c | c c|c c|}
\hline
\hline
HD~49933 & & Scenario 1  &&Scenario 2&\\
\hline
Degree & Order& Frequency & Uncertainty&Frequency & Uncertainty \\
& & $\mu$Hz    &$\mu$Hz &  $\mu$Hz    &$\mu$Hz\\
 \hline
   0 &    14 &       1244.7 &      0.3 &       1288.2 &      0.9 \\
   0 &    15 &       1329.6 &      0.2 &       1372.4 &      0.6 \\
   0 &    16 &       1415.8 &      0.5 &       1455.3 &      0.5 \\
   0 &    17 &       1501.6 &      0.3 &       1545.3 &      0.6 \\
   0 &    18 &       1586.8 &      0.4 &       1630.4 &      1.2 \\
   0 &    19 &       1670.4 &      0.7 &       1717.3 &      0.8 \\
   0 &    20 &       1756.8 &      0.9 &       1802.1 &      1.2 \\
   0 &    21 &       1840.6 &      1.1 &       1882.3 &      0.8 \\
   0 &    22 &       1926.6 &      1.6 &       1974.1 &      0.8 \\
   0 &    23 &       2013.7 &      1.1 &       2062.1 &      1.1 \\
   0 &    24 &       2100.8 &      1.0 &       2147.3 &      0.9 \\
   0 &    25 &       2185.9 &      2.0 &       2236.0 &      1.8 \\
   0 &    26 &       2279.8 &      1.0 &       2324.5 &      1.4 \\
   0 &    27 &       2362.9 &      3.1 &       2403.7 &      2.8 \\
    \hline
   1 &    14 &       1289.9 &      0.3 &       1244.3 &      0.6 \\
   1 &    15 &       1372.3 &      0.1 &       1329.3 &      0.8 \\
   1 &    16 &       1458.3 &      0.2 &       1415.5 &      0.5 \\
   1 &    17 &       1545.2 &      0.3 &       1501.5 &      0.5 \\
   1 &    18 &       1629.7 &      0.4 &       1586.2 &      0.6 \\
   1 &    19 &       1714.2 &      0.4 &       1670.3 &      0.7 \\
   1 &    20 &       1800.3 &      1.0 &       1756.6 &      0.8 \\
   1 &    21 &       1885.0 &      0.9 &       1840.5 &      0.8 \\
   1 &    22 &       1973.1 &      0.5 &       1927.1 &      1.4 \\
   1 &    23 &       2058.7 &      0.7 &       2013.6 &      0.9 \\
   1 &    24 &       2144.5 &      1.2 &       2101.5 &      1.3 \\
   1 &    25 &       2233.0 &      1.0 &       2188.7 &      3.5 \\
   1 &    26 &       2321.5 &      2.4 &       2279.1 &      1.1 \\
   1 &    27 &       2403.9 &      1.7 &       2362.1 &      2.3 \\
    \hline
   2 &    14 &       1240.2 &      0.2 &       1298.1 &      1.0 \\
   2 &    15 &       1331.1 &      0.1 &       1369.0 &      0.7 \\
   2 &    16 &       1410.7 &      1.0 &       1460.7 &      1.0 \\
   2 &    17 &       1499.0 &      0.7 &       1535.4 &      1.0 \\
   2 &    18 &       1587.7 &      0.8 &       1630.2 &      2.5 \\
   2 &    19 &       1667.8 &      1.0 &       1712.0 &      1.3 \\
   2 &    20 &       1755.8 &      1.3 &       1796.3 &      2.4 \\
   2 &    21 &       1840.3 &      0.9 &       1887.6 &      1.4 \\
   2 &    22 &       1926.0 &      0.7 &       1968.5 &      3.6 \\
   2 &    23 &       2016.6 &      1.8 &       2055.5 &      1.3 \\
   2 &    24 &       2103.0 &      1.6 &       2141.5 &      1.4 \\
   2 &    25 &       2190.4 &      1.1 &       2230.2 &      1.8 \\
   2 &    26 &       2271.1 &      2.6 &       2314.5 &      2.3 \\
   2 &    27 &       2362.3 &      3.5 &       2403.7 &      5.0 \\

\hline    
\hline
Order&& Width & Uncertainty&Width &Uncertainty \\
&& $\mu$Hz    &$\mu$Hz &  $\mu$Hz  &  $\mu$Hz  \\
 \hline
   14 &  &         1.03 & +0.52/-0.35 &         3.31 & +1.37/-0.97 \\
   15 &  &         0.53 & +0.34/-0.21 &         1.57 & +1.58/-0.79 \\
   16 &  &         1.29 & +0.33/-0.26 &         2.38 & +0.82/-0.61 \\
   17 &  &         1.13 & +0.41/-0.30 &         2.52 & +1.17/-0.80 \\
   18 &  &         2.15 & +0.52/-0.42 &         4.14 & +1.08/-0.86 \\
   19 &  &         2.69 & +0.64/-0.52 &         4.79 & +1.20/-0.96 \\
   20 &  &         4.56 & +0.92/-0.77 &         7.18 & +1.39/-1.16 \\
   21 &  &         3.18 & +0.65/-0.54 &         4.86 & +1.03/-0.85 \\
   22 &  &         2.63 & +0.87/-0.65 &         4.91 & +1.18/-0.95 \\
   23 &  &         3.63 & +1.09/-0.84 &         5.52 & +1.24/-1.01 \\
   24 &  &         3.47 & +1.41/-1.00 &         5.27 & +1.63/-1.24 \\
   25 &  &         3.63 & +1.59/-1.11 &         6.31 & +1.83/-1.42 \\
   26 &  &         5.93 & +2.47/-1.74 &         7.46 & +2.38/-1.81 \\
   27 &  &         6.28 & +3.27/-2.15 &         9.17 & +2.79/-2.14 \\
   \hline
\hline
Height&$H\ind{H}$ (ppm$^2\mu$Hz$^{-1}$) &1.28 & +0.26/-0.22 &   0.69 & +0.14/-0.12\\
&$\nu\ind{H}$ ($\mu$Hz) &1457.9 & 112.9 & 1589.1 & 88.1\\
&$\sigma\ind{H}$ ($\mu$Hz) &655.8 & 83.2 &  625.9 & 76.8\\
\hline
Background&B  (ppm$^2\mu$Hz$^{-1}$)&0.14 & +0.01/-0.01 &         0.14 & +0.01/-0.01 \\
&K (ppm$^2\mu$Hz$^{-1}$)&0.77 & +0.03/-0.03 &         0.75 & +0.03/-0.03 \\
&C ($10^{-6}$s$^2$)&1.33 & +0.06/-0.06 &         1.36 & +0.07/-0.06 \\
\hline
Splitting&$\nu\ind{S}$ ($\mu$Hz) &3.3 &  0.1 &    3.3 &  0.4\\
Inclination &$i$ (deg)& 52 &    4 &     -8 &   28\\
Likelihood&$l\ind{MAP}$&54.8 & &   59.9&\\
\hline    
\end{tabular}
\label{tab:campagne}
\end{table*}

\bibliographystyle{aa} 
\bibliography{bibi} 

\end{document}